\documentclass{pas}

\usepackage{aas-macros}
\usepackage{xcolor}
\usepackage{hyperref}

\RequirePackage{etex}

\newcommand{\hi}{H {\sc i} }

\newcommand{\OI}{\mbox{[O\,\textsc{i}]}}
\newcommand{\NII}{\mbox{[N\,\textsc{ii}]}}

\usepackage{multirow}
\usepackage{graphicx}

\begin{document}
	
	\lefttitle{Publications of the Astronomical Society of Australia}
	\righttitle{Tamhane et. al.}
	
	\jnlPage{1}{18}
	\jnlDoiYr{2026}
	\doival{10.1017/pasa.xxxx.xx}
	
	\articletitt{Research Paper}
	
	\title{HST view of NGC 5044: Constraints on Filament Widths, Magnetic Support, Multiphase Structure, and Comparison with Cluster Environments}
	
	\author{\sn{Prathamesh} \gn{Tamhane}$^{1}$,
				 \sn{Ming} \gn{Sun}$^{1}$,
			     \gn{William} \sn{Waldron}$^{1}$,
			     \gn{Kokoro} \sn{Hosogi}$^{1}$,
			     \gn{Patricia} \sn{da Silva}$^{1}$,
			     \gn{Huan} \sn{Le}$^{1}$,\\
			     \gn{Massimo} \sn{Gaspari}$^{2}$,
			     \gn{Francoise} \sn{Combes}$^{3}$,
			     \gn{Norbert} \sn{Werner}$^{4}$,
			     \gn{Gerrit} \sn{Schellenberger}$^{5}$,
			     \gn{Andrew} \sn{Fabian}$^{6}$,\\
			     \gn{Rebecca} \sn{Canning}$^{7}$,
			     \gn{Laurence} \sn{David}$^{5}$,
			     \gn{Megan} \sn{Donahue}$^{8}$ and
			     \gn{Mark} \sn{Voit}$^{8}$
			    }
	
	\affil{$^1$Department of Physics and Astronomy, University of Alabama in Huntsville, 301 Sparkman Drive, Huntsville, AL 35899, USA\\
	        $^2$Department of Physics, Informatics and Mathematics, University of Modena and Reggio Emilia, 41125 Modena, Italy\\
            $^3$LUX, Observatoire de Paris, PSL Univ., Coll\`{e}ge de France, CNRS, Sorbonne Univ., Paris, 75014, France\\
            $^4$Department of Theoretical Physics and Astrophysics, Faculty of Science, Masaryk University, Kotl\'a\v{r}sk\'a 2, Brno, 611 37, Czech Republic\\
            $^5$Center for Astrophysics \textbar{} Harvard \& Smithsonian, 60 Garden Street, Cambridge, MA 02138, USA\\
            $^6$Institute of Astronomy, Madingley Road, Cambridge CB3 0HA, UK\\
            $^7$Institute of Cosmology \& Gravitation, University of Portsmouth, Dennis Sciama Building, Portsmouth, PO1 3FX, UK\\
            $^8$Michigan State University, Physics and Astronomy Department, East Lansing, MI 48824-2320, USA
        }
	
	\corresp{P. Tamhane, M. Sun, Email: \href{mailto:pdt0003@uah.edu}{pdt0003@uah.edu}, \href{mailto:ms0071@uah.edu}{ms0071@uah.edu}} 
	
	
	\history{(Received 3 October 2025; revised 3, 20 and 25 February 2026; accepted 27 February 2026)}
	
	\begin{abstract}
		We present new Hubble Space Telescope (HST) imaging of the ionised filaments in the brightest group galaxy NGC 5044, providing the first high-resolution view of such structures in a galaxy group. The filaments extend several kiloparsecs from the centre, with widths of $\sim$50--120 pc. Some strands are as narrow as those in cluster cores, while others are broader, consistent with the weaker confining pressure of the intragroup medium. With our limited sample, we find that the filament width ($W$) roughly scales with ambient pressure ($P$) as $W \propto P^{-0.4}$.
		
		Combining HST with molecular and MUSE observations, we measure column densities and magnetic field strengths. Equipartition magnetic fields decline from $\sim$40 $\mu$G near the centre to $\sim$20 $\mu$G at 5 kpc, about 2--3 times weaker than in clusters. Dynamical stability arguments require stronger radial magnetic fields ($\sim$10$^2$ $\mu$G), consistent with simulations and magnetic field lines draping and flux freezing around cavities, though such high values may be difficult to reconcile with Faraday Rotation Measure limits. Turbulence and cosmic rays can also provide complementary support.
		
		Filaments are stable against gravitational collapse, and ultraviolet imaging reveals no star formation in NGC 5044 ($<$10$^{-3}$ M$_\odot$ yr$^{-1}$), confirming that star formation in filaments in both groups and clusters remains largely quenched. NGC 5044 hosts an ionised gas core within its Bondi radius with $n_e \propto r^{-1}$ and filling factor $f \gtrsim 3 \times 10^{-3}$, that is connected to the extended filaments, suggesting a channel for gas inflow toward the black hole.
		
		Our results show that group filaments share the same origin and stabilising mechanisms as cluster filaments, with magnetic fields and AGN feedback preserving filamentary structures with ambient pressure and dust survival as key factors for molecular gas formation and survival. Lower pressure groups favour broader, diffuse filaments with sporadic molecular clumps and less dust shielding, while higher pressure clusters host narrower strands with stronger molecular/ionised gas alignment. We predict that (i) filament widths scale with ambient pressure, (ii) filament-coincident Faraday rotation structures should appear at $\leq$0.1 kpc resolution, and (iii) molecular/ionised gas co-spatiality is weaker in groups than in clusters.
	\end{abstract}
	
	\begin{keywords}
		Active galaxies; Interstellar filaments; Interstellar magnetic fields; Cluster of galaxies; Groups of galaxies; Active galactic nuclei
	\end{keywords}
	
	\maketitle
	
	\section{INTRODUCTION }
	\label{sec:int}
	
	Over one-third of galaxy clusters and groups host dense X-ray emitting cores centred on their brightest central galaxies (BCGs), with radiative cooling times shorter than 1 Gyr at radii $\lesssim$ 20 kpc \citep[e.g.,][]{mcnamara12,donahue22}. In the absence of heating, these `cool cores' should drive substantial cooling flows, leading to the rapid accumulation of cold gas at rates up to $\sim$1000 M$_\odot$ yr$^{-1}$ \citep[e.g.,][]{fabian94}. However, sensitive X-ray spectroscopic observations have not detected the expected reservoirs of gas cooling through soft X-ray lines, indicating that most of the radiative cooling must be offset by heating processes—primarily active galactic nucleus (AGN) feedback, as well as thermal conduction and turbulence \citep[e.g.,][]{fabian12,donahue22}. Yet this heating cannot perfectly balance cooling at all times and spatial scales. Multi-wavelength observations have now established that cold gas and young stars are frequently present in cool cores, albeit at reduced levels compared to classical cooling flow predictions. These include CO emission \citep[e.g.,][]{edge01,salome03}, cold atomic gas and dust \citep[e.g.,][]{edge10,rawle12}, warm molecular hydrogen \citep[e.g.,][]{donahue11}, H$\alpha$ nebulae \citep{crawford99,hamer16}, and young star clusters \citep[e.g.,][]{mcdonald11,canning10,canning14}.
	
	One of the most striking features of cool cores is the extended, filamentary optical line-emitting nebulae. Perhaps the best example is NGC~1275 \citep[the BCG of the Perseus cluster, e.g.,][]{crawford99,fabian08}. These narrow, long-lived filaments trace the presence of multiphase gas and are often co-spatial with cold molecular clouds. Their low-ionisation spectra with elevated \NII{}/H$\alpha$ and strong \OI{} lines suggest distributed, non-stellar ionisation sources such as heat conduction, turbulent mixing, cosmic rays, or Magneto-Hydro-Dynamic (MHD) wave dissipation \citep[e.g.,][]{ferland09,canning16}. Regardless of the specific mechanism, the optical filaments are sites of condensation from the hot intracluster medium (ICM), and thus offer critical insights into the physical conditions and thermodynamic history of cool cores.
	
	The optical emission-line nebulae observed in cool cores are widely interpreted as signatures of multiphase gas condensation arising from thermal instabilities in a hot ambient medium \citep[e.g.,][]{mccourt12,gaspari12}. Heating from AGN, stellar winds, and supernovae offset cooling globally, but they are often unable to prevent localised overdensities from becoming thermally unstable. These overdensities may develop along turbulent eddies \citep{gaspari17} or within the low-entropy gas uplifted by buoyant bubbles and jets launched by the central AGN \citep{revaz08}. As gas is displaced to larger radii, it can cool and condense into warm and cold filaments, a process described by precipitation and uplift-driven condensation models \citep[e.g.,][]{mcnamara16,voit17}. Some of this condensed material may subsequently accrete onto the supermassive black hole, potentially in a disordered manner as described by chaotic cold accretion \citep[CCA,][]{pizzolato05,gaspari13}. Thus, the optical emission-line nebulae provide important kinematic and timescale constraints on gas flows in cool cores, from $\sim$ 50 kpc down to regions within the Bondi radius on scales of 50 -- 100 pc \citep{fabian08,fabian16,hamer16,olivares19}.
	
	\begin{figure*}
		\centering
		\includegraphics[width=\textwidth]{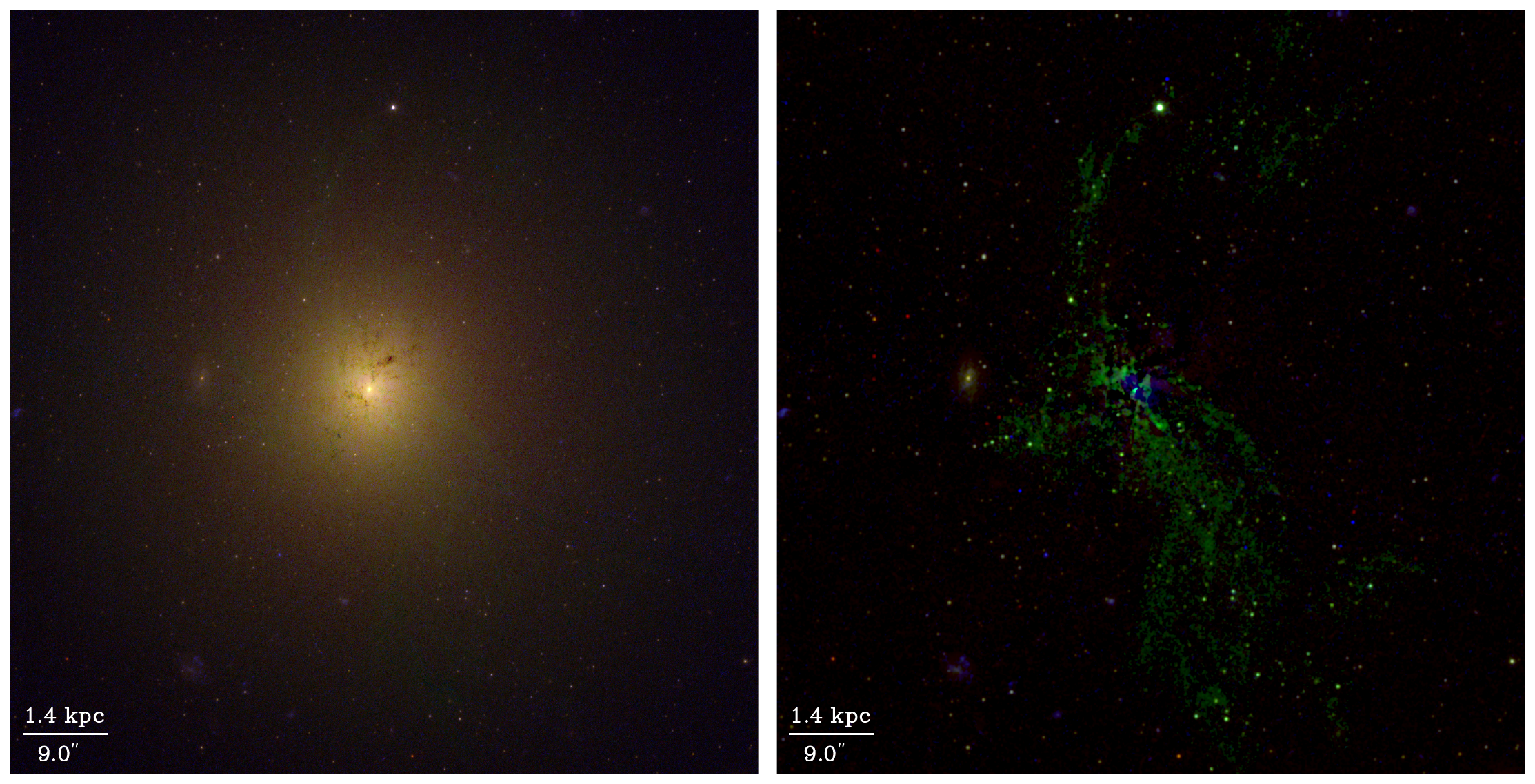}
		\caption{{\em Left:} RGB composite of NGC~5044 combining HST F300X (blue), F665N (green), and F814W (red) images. The F665N filter is centred on the H$\alpha$ line. The images are modified to enhance the dust filaments in the left panel and H$\alpha$ filaments in the right panel. {\em Right:} Same image after subtracting the galactic continuum from each filter before combining, highlighting the inner filaments and showing no young blue star clusters.
		}
		\label{fig:ngc5044_rgb}
	\end{figure*}
	
	While much effort has gone to the study of cluster cool cores, group cool cores are also important as baryon physics (e.g., preheating and AGN feedback) begins to dominate over gravity in these low-mass halos. To explore this regime, we have initiated several multi-wavelength studies of group cool cores, including an H$\alpha$ imaging survey targeting the brightest group galaxies (BGGs) in over 60 nearby systems using the 4.1-m Southern Astrophysical Research  Telescope (SOAR) and 3.5-m Apache Point Observatory (APO) telescopes (PI: Sun). These observations reveal that extended warm filaments, often seen in clusters, are also common in group environments, though their physical conditions and formation mechanisms remain less well understood \citep{Sun12,Lakhchaura18}
	
	As a key case study, we focus in this paper on NGC~5044, one of the nearest and best studied cool core groups. NGC~5044 hosts one of the brightest and most extended optical emission-line nebulae in the sample \citep{Sun12,David14,werner14}. It is then naturally an ideal system for detailed studies.
	The radio luminosity of NGC 5044 \citep[$\sim 5 \times 10^{21}$ W Hz$^{-1}$;][]{grossova22} is 2000 times weaker than that of NGC 1275 \citep[$\sim 10^{25}$ W Hz$^{-1}$;][]{pedlar90} at $\sim$1.4 GHz, which may imply a substantially smaller energy density of cosmic rays. It also hosts few $10^7$ M$_\odot$ of molecular gas \citep{David14,temi18} as well atomic gas detected in [C {\sc ii}] \citep{werner14} and from \hi absorption \citep{rajpurohit25}. Molecular clouds have also been detected in NGC 5044 in absorption indicating an inflow of gas \citep{schellenberger20,rose23}. We obtained Hubble Space Telescope (HST) imaging of NGC 5044 providing a uniquely sharp view of its optical filaments, enabling measurements of their morphology to constrain the strength of the magnetic field \citep[e.g.,][]{fabian08}. Combined with supporting observations from Atacama Large Millimeter/submillimeter Array (ALMA), Chandra, and Multi Unit Spectroscopic Explorer (MUSE) spectrograph on the Very Large Telescope (VLT), as well as comparisons to the latest MHD simulations, these high-resolution observations offer a comprehensive test of magnetically supported condensation and accretion in a group cool core. We compared filament properties in NGC 5044 with those in M87, Centaurus and Perseus cluster filaments to examine whether the filament properties in group environments differ systematically from their more massive cluster counterparts, and assess the role of non-thermal processes in shaping the multiphase structure at the low-mass end of the halo mass function.
	At the distance of NGC 5044 of 31.2 Mpc \citep{tonry01}, 1$'' = 151$ pc. For Centaurus, Perseus, and M87, assuming distances of 39.8 Mpc, 76.7 Mpc, and 16.5 Mpc, the corresponding spatial scales are 1$''$ = 190, 359, and 80 pc, respectively. We used $cz = 2747$ km s$^{-1}$ for the system velocity of NGC 5044.
	
	\section{DATA ANALYSIS}
	\label{sec:data_analysis}
	
	\subsection{HST}
	\label{sec:da_hst}
	
	The HST data for NGC 5044 used in this work came from proposal ID 15290 (PI: Sun) and were collected with Wide Field Camera 3 (WFC3) instruments from F300X, F665N, and F814W filters. The F300X reveals young stellar populations if present, the F665N filter reveals the optical emission-line nebula, and the F814W filter shows the evolved stellar population. Table~\ref{tab:obs} summarises the HST data used in this analysis.
	
	The images were aligned to the GAIA~Data Release 3
	catalogue \citep{gaia2016, gaia2023} to within 0$''$.01 using the
	\textsc{TweakReg} tool from the \textsc{DrizzlePac} package. 
	For data with a small field of view,
	the HST pipeline drizzled images from each dither position were used in alignment, and the final alignment was back-propagated to the input images using \textsc{TweakBack} within the \textsc{DrizzlePac}.
	Images in the same band taken on different dates were photometrically normalised using
	\textsc{PhotEq} in the \textsc{DrizzlePac}.
	Subsequently, the images in each band were drizzled together using \textsc{AstroDrizzle},
	which reduced noise, removed cosmic rays in overlapping regions, and produced a
	consistent pixel scale of 0$''$.03 for all bands.
	All the mosaic images related to this work will be released to the Mikulski Archive for Space Telescopes as High Level Science Products.\footnote{\url{https://mast.stsci.edu/hlsp/\#/}}. An RGB image of NGC 5044 is shown in Figure~\ref{fig:ngc5044_rgb}. 
	
	We constructed models of diffuse galactic emission from the F665N and F814W images for NGC 5044 using isophote fitting and subtraction with \textsc{Photutils}\footnote{\url{https://photutils.readthedocs.io/en/stable/}}. These models were subtracted from the data to obtain residuals highlighting low-surface brightness filaments. Since this method also removes a significant component of the ionised gas core in F665N image, we scaled the F814W model and its isophotes to match the isophotes of F665N derived using the same method between 20 and 500 pixel radius region where both isophotes have the same shape. Subtracting this scaled model from F665 original image produced a residual map highlighting the extended H$\alpha$+[N\,\textsc{ii}] filaments and the central core. Since the underlying stellar continuum should not differ significantly between the F814W and F665N images in the central regions, both methods produce consistent filament morphologies.
	
	\begin{table*}
		\begin{center}
			\caption{HST data used in this study from the proposal 15290 (PI: Sun). Note: $^{\rm a}$: the number of exposures.}
			\label{tab:obs}
			\begin{tabular}{ccccccc}
				\hline
				Filter & Instrument & Obs Date & N$^{\rm a}$ & Total Exposure (sec) & Mean $\lambda$ (\AA) & Area (arcmin$^2$) \\
				\hline
				F300X  & WFC3/UVIS    & 2018-07-28 & 3 & 2607  & 2882 & 7.5 \\
				F665N  & WFC3/UVIS    & 2018-07-28, 2018-08-19 & 8 & 10128 & 6561 & 7.5 \\
				F814W  & WFC3/UVIS    & 2018-07-28 & 2 & 834 & 8117 & 7.5 \\
				\hline
			\end{tabular}
		\end{center}
	\end{table*}
	
\begin{figure*}
		\includegraphics[width=\textwidth]{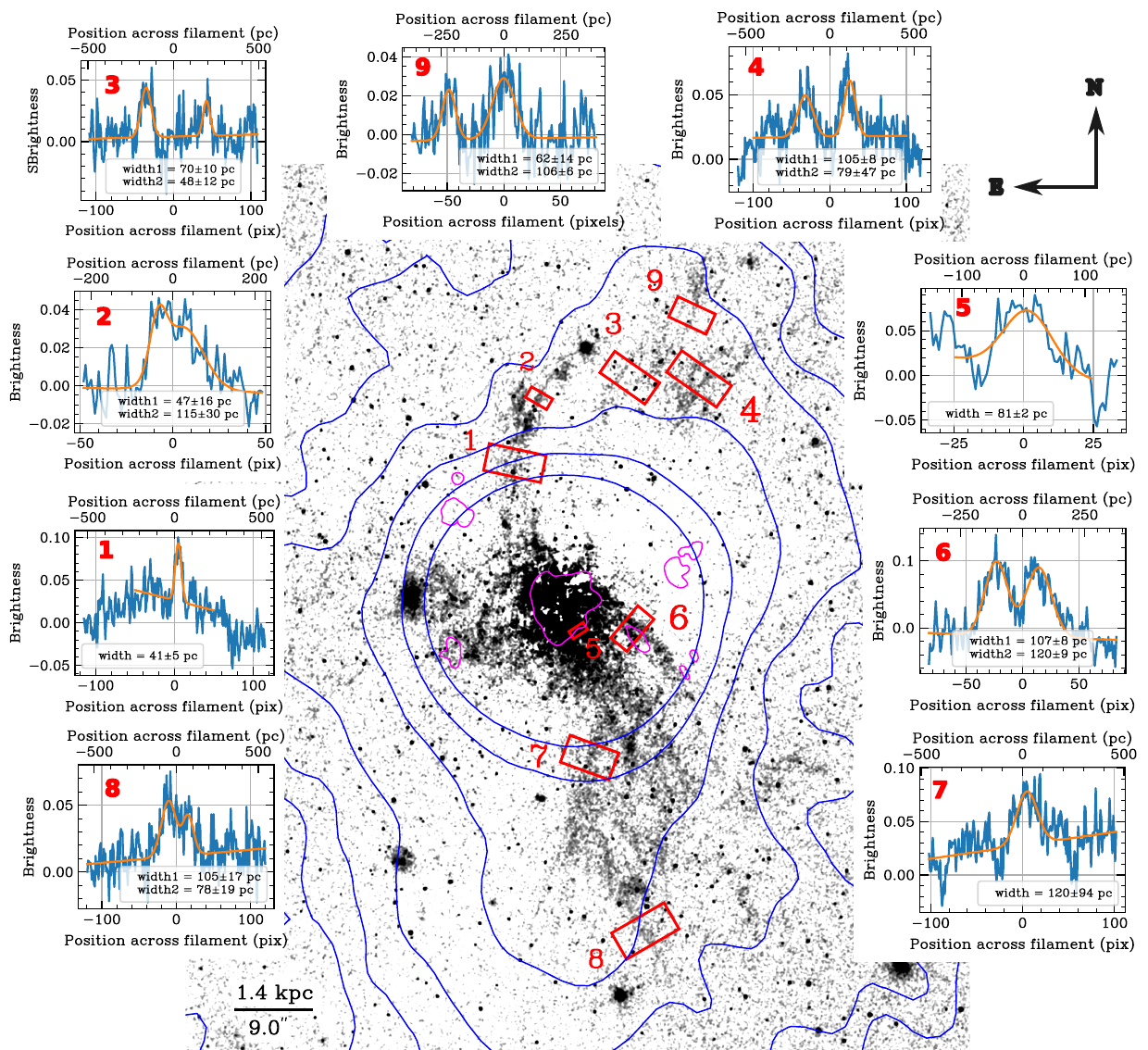}
		\caption{F665N residual image showing extended H$\alpha$+[N\,\textsc{ii}] filaments in NGC 5044. The image is smoothed to enhance the visibility of the filaments. Red rectangles mark the regions where filament surface-brightness profiles were extracted in the unsmoothed image. The extracted profiles (blue) and their Gaussian fits (orange) are shown in the accompanying panels. We also show the ALMA CO(2-1) contours based in the analysis presented in \citet{tamhane22} in magenta. The Giant Metrewave Radio Telescope 380 MHz contours from \citep{rajpurohit25} are shown in blue. The magenta regions show CO contours at 0.17 Jy km s$^{-1}$ in moment 0 maps which corresponds to 2.5-$\sigma$ detection.
		}
		\label{fig:N5044}
	\end{figure*}

	We selected rectangular regions centred on filaments in various parts of the galaxy, as shown in Figure~\ref{fig:N5044}, and extracted surface brightness profiles along the axis perpendicular to the filaments. Before extracting the profiles, we masked bright point sources using \textsc{photutils} segmentation, applying a threshold of 1.5$\sigma$ per pixel above the background RMS level and 2D Gaussian kernel with a full width at half maximum (FWHM) of 3 pixels, to minimise contamination. We then fitted one or two Gaussian models to these profiles and measured their FWHM, which we adopted as the characteristic width of the filaments for our analysis. We estimated filament lengths by visually tracing the projected extent of filaments along their major axis over which the coherent emission was detected in the image. We note that the widths are not constant along the filaments, often varying by up to 50\%, but our Gaussian fits to surface brightness profiles provide representative values.
	
\begin{figure*}
		\includegraphics[width=\textwidth]{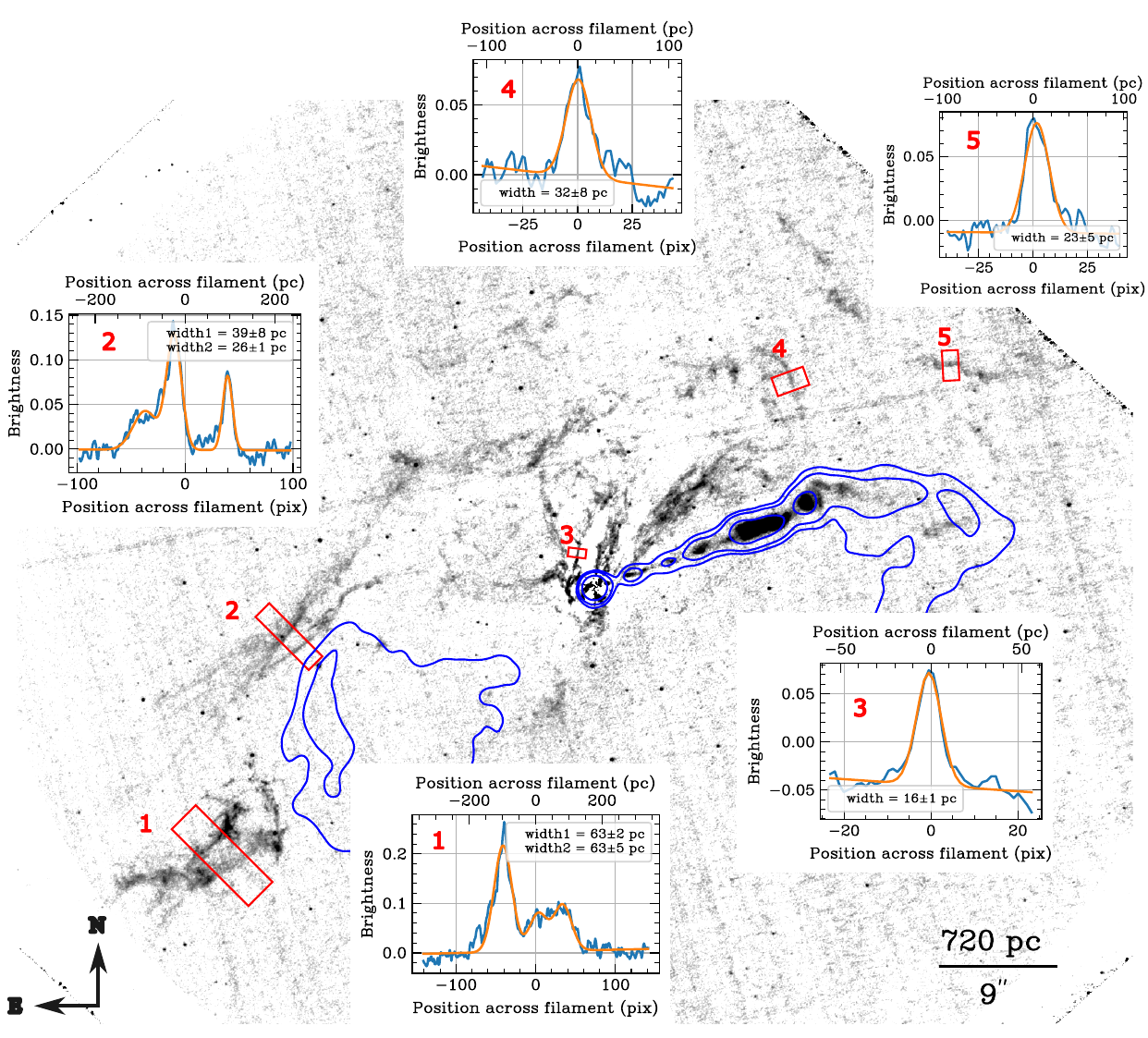}
		\caption{F660N-Pol residual image showing extended H$\alpha$+[N\,\textsc{ii}] filaments in M87. The image is smoothed to enhance the visibility of the filaments. Red rectangles mark the regions where filament surface-brightness profiles were extracted in the unsmoothed image. The extracted profiles (blue) and their Gaussian fits (orange) are shown in the accompanying panels. The Very Large Array L band radio contours are shown in blue.}
		\label{fig:m87}
	\end{figure*}

\begin{figure*}
		\includegraphics[width=\textwidth]{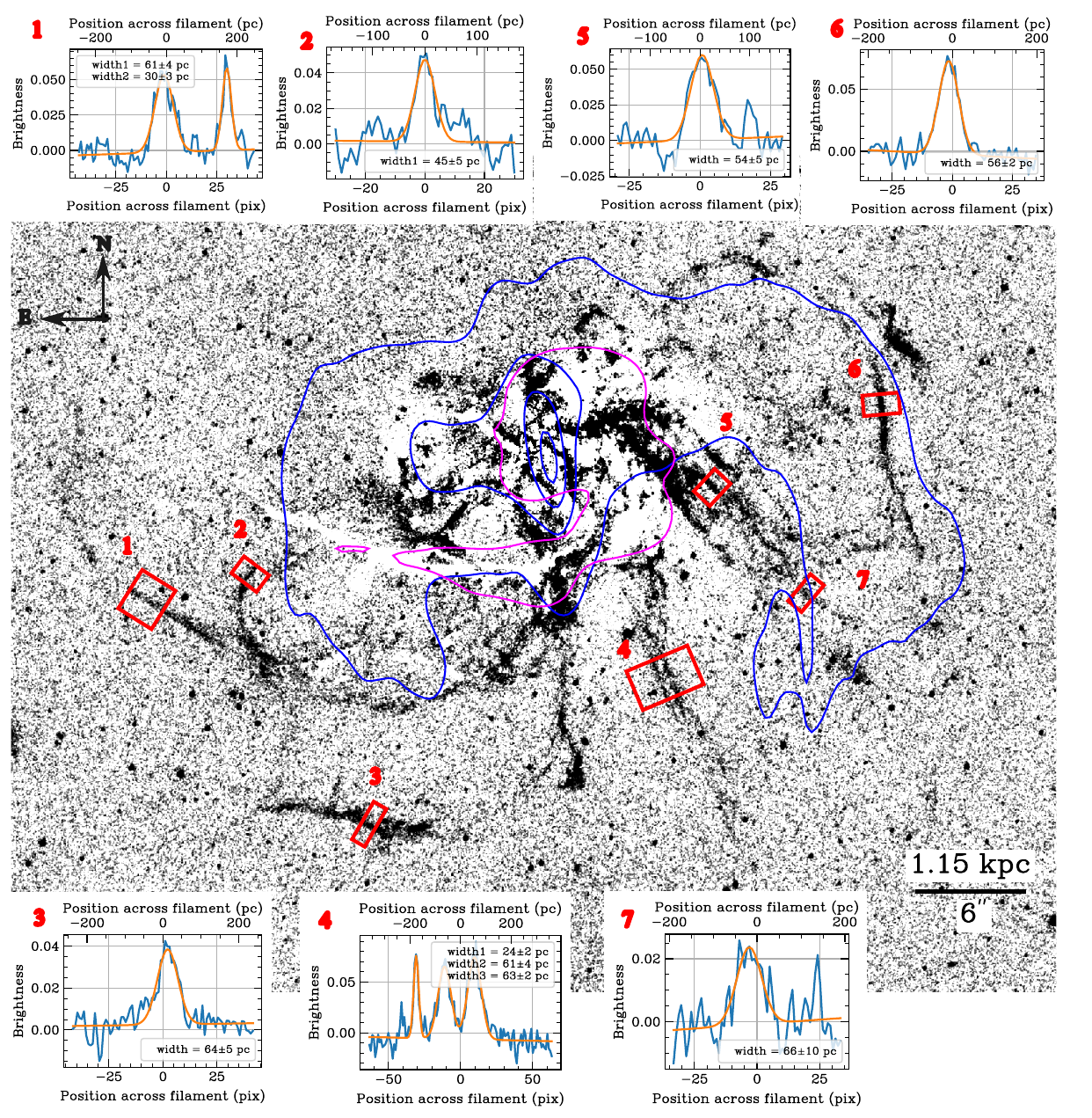}
		\caption{F665N residual image showing extended H$\alpha$+[N\,\textsc{ii}] filaments in NGC4696. The image is smoothed to enhance the visibility of the filaments. Red rectangles mark the regions where filament surface-brightness profiles were extracted in the unsmoothed image. The extracted profiles (blue) and their Gaussian fits (orange) are shown in the accompanying panels. Additional narrow peaks in panels 1 and 4 are point sources. Magenta contour shows the ALMA CO(1-0) detection in moment 0 map of \citet{tamhane22} at 0.06 Jy km s$^{-1}$. The Very Large Array L band radio contours are shown in blue.}
		\label{fig:ngc4696}
	\end{figure*}

	We re-analysed HST WFC3/UVIS F665N data using a similar data-analysis procedure for the Centaurus, with observations from Proposal ID 13308 (PI: Fabian), and subtracted the galactic continuum from the final image with \textsc{photutils} to reveal the filamentary structure (Figure~\ref{fig:ngc4696}). For M87, we used the HST ACS/WFC F660N-Pol image presented in \citet{tamhane25}, to analyse the nebular filaments (Figure~\ref{fig:m87}).
	
	\subsection{MUSE}
	\label{sec:da_muse}
\begin{figure*}
		\centering
		\includegraphics[width=\textwidth]{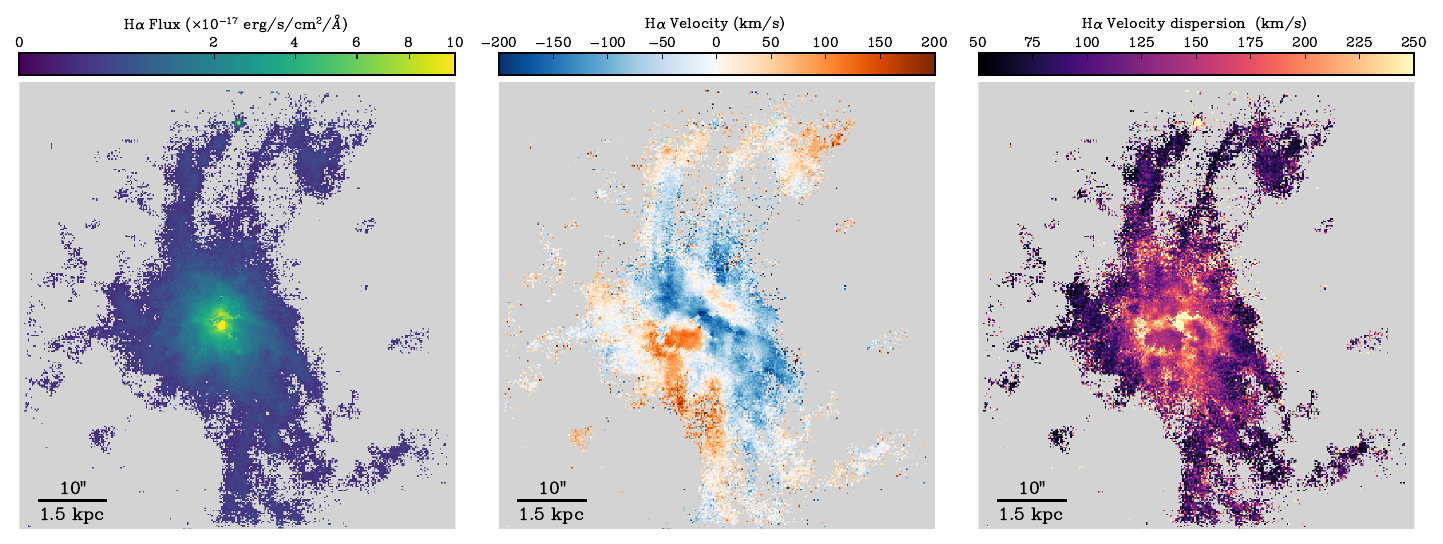}
		\vspace{-0.7cm}
		\caption{Maps of H$\alpha$ flux (left), velocity (centre) and velocity dispersion (right) generated from MUSE datacube. North is up and East is to the left.
		}
		\label{fig:N5044_musemaps}
	\end{figure*}

	We also used data of NGC 5044 obtained with the MUSE spectrograph on the VLT in the Wide Field Mode (WFM), observed as part of program 094.A-0859 (PI: Hamer). The galaxy was observed in 2015 through five exposures, with a total integration time of 10.66 ks, covering the wavelength range from $\sim$4800 to 9300~\AA. The data have an average spatial resolution of 0.$''$7 and a spectral resolution of $R\sim3000$ (corresponding to FWHM $\sim$100 km s$^{-1}$ at 7000~\AA). The MUSE data were reduced using the latest version (2.8.5) of the MUSE pipeline \citep{weilbacher14} and the EsoRex command-line tool. We then fitted the resulting data cube using the TARDIS data analysis pipeline\footnote{\url{https://gitlab.com/francbelf/ifu-pipeline}}, a Python-based spectral fitting program that incorporates the penalised pixel-fitting method (pPXF), based on Physics at High Angular resolution in Nearby GalaxieS (PHANGS)-MUSE Data Analysis Pipeline (DAP) \citep{emsellem22}. The TARDIS emission-line maps were used in this analysis to derive the H$\alpha$ flux and velocity distributions in the galaxy. The H$\alpha$ flux, velocity and velocity dispersion maps derived from this data are shown in Figure~\ref{fig:N5044_musemaps}. We derived the radial velocity of the galaxy based on the average velocity of the stars within the 1.5$''$ radius region around the nucleus and found a velocity of $cz = 2747$ km s$^{-1}$. We use this velocity to make the velocity maps in this paper. We note that this value is close to $cz = 2757$ km s$^{-1}$ used by \citet{rajpurohit25}.
	
	\section{RESULTS}
	\label{sec:results}
	
	\subsection{Filament Morphologies}
	In NGC 5044, M87 and Centaurus, we measured filament morphologies at several locations along the length. We extracted surface brightness profiles of filaments in rectangular regions placed along portions of the filaments, where the local morphology is approximately linear over several resolution elements giving a well-defined transverse profile and the region contains clearly identifiable narrow strands, enabling us to assess how thin the filamentary structures can become at HST resolution. These criteria ensured that the measured widths reflect intrinsic filament structure rather than artifacts of curvature or noise. We then fit a gaussian profiles to the extracted profiles to estimate the full width at half maximum (FWHM) of filament. The extraction regions, surface brightness profiles and the corresponding fits are shown in Figures~\ref{fig:N5044},~\ref{fig:m87} and~\ref{fig:ngc4696}.
	
	\subsubsection{NGC 5044}
	Figure~\ref{fig:N5044} shows H$\alpha$+[N\,\textsc{ii}] filaments extending asymmetrically from the central regions of the galaxy. The brightest emission originates from the central 1.5 kpc from the nucleus, likely tracing a dense ionised gas component. The filaments extend $\sim$7 kpc to the north and $\sim$5.8 kpc to the south of the nucleus. The northern filaments show more substructure compared to the southern filaments, which appear thicker and have more bends.  Interestingly, only the northern filament, specifically in regions 1 and 2, is co-spatial with the X-ray filament identified by \citet{david17}. The average width of the southern filaments is approximately 115 pc, while the northern filaments are thinner, with widths ranging from 40 to 80 pc with some filaments with widths more than 100 pc. Overall, the characteristic filament widths range from 50 to 120 pc, as shown in the brightness profile panels in Figure~\ref{fig:N5044}. The average filament width is $\sim$85 pc. In both regions the individual filament lengths vary from 1 to 3 kpc. Thus, while most filaments in NGC 5044 appear broader than those typically seen in cluster environments, some filaments have narrow widths comparable to cluster filaments despite being in a stratified environment suggesting highly magnetised filaments.
	
	The MUSE ionised gas flux, velocity, and velocity dispersion maps for NGC~5044 are shown in Figure~\ref{fig:N5044_musemaps}. The ionised gas is spatially coincident with the narrow filaments identified in the HST imaging. The velocity map reveals coherent and smooth velocity gradients along all major filaments. The northern filaments show a smooth velocity gradient with line-of-sight velocities varying from $\sim$40 km s$^{-1}$ at the projected outer ends to $\sim -165$ km s$^{-1}$ toward the central regions. The southern filament shows an opposite velocity gradient, increasing from $\sim 0$ km s$^{-1}$ at larger projected radii to $\sim 135$ km s$^{-1}$ closer to the centre. Such velocity gradients are expected if gas uplifted by radio jets is subsequently falling back toward the galaxy centre. As described in Section~\ref{sec:dust}, the southern filament lies on the near side of the galaxy, for which the observed velocity gradient is consistent with infalling gas. The velocity dispersion map shows relatively uniform dispersions of $\sim$80--100 km s$^{-1}$ along the filaments. In contrast, the velocity dispersion increases significantly to $\sim$350 km s$^{-1}$ in regions close to the nucleus in projection, which may indicate disturbed gas motions or multiple kinematic components along the line of sight. The nuclear region has very high velocity dispersion exceeding 500 km s$^{-1}$ with multiple velocity components, consistent with the results of \citet{diniz17}, who reported an FWHM of 3000 km s$^{-1}$ in the nucleus and suggested the presence of an outflow based on Gemini Multi-Object Spectrograph data. The detailed analysis of the MUSE data will be presented in the future paper. Overall, the velocity field suggests slow inflow motions. These low velocities are consistent with the standard picture of the cooling of the infalling gas, also seen in other such similar systems \citep[e.g.,][]{olivares19}.
	
	Most of the molecular gas in NGC 5044 is concentrated within the central 2 kpc, with only a few giant molecular associations (GMAs) scattered around the centre, as shown by magenta contours in Figure~\ref{fig:N5044}. In general, these GMAs are not co-spatial with the extended filaments, except in region 6, where a GMA coincides with the filament in projection. ACA observations did not detect molecular gas outside the inner 2 kpc \citep{schellenberger20}, although Herschel [C \textsc{ii}] $\lambda$158 $\mu$m emission extends to 8 kpc \citep{werner14}. Given the lack of significant star formation, this emission likely traces cold atomic gas, suggesting that a more extended, diffuse molecular component may exist below the detection threshold of current observations.
	
	This trend is also reflected in the H$\alpha$ surface brightness. In the MUSE map, the average H$\alpha$ surface brightness within the central $\sim$0.7 kpc radius region where CO is detected in NGC~5044 is 10--15 times higher than in the extended filaments. In contrast, in region~6 (Figure~\ref{fig:N5044}), where both H$\alpha$ and CO are detected along a filament, the H$\alpha$ surface brightness is $\sim$4 times lower than in the central region, while the CO surface flux density is a factor of $\sim$2 lower. Given the substantially lower H$\alpha$ surface brightness of the extended filaments, the non-detection of CO in these regions is consistent with the expected low molecular surface densities and current observational sensitivity limits. Nevertheless, the partial co-spatiality of molecular and ionised gas indicates that molecular clouds can sometimes condense within filaments, but in most regions, the cold phase appears decoupled from the ionised gas, reflecting local variations in pressure, density, or feedback.
	
	The morphological differences between the northern and southern filaments in NGC 5044 may reflect different evolutionary stages in the cooling process and their lifetimes. As suggested by \citet{david17}, not all X-ray filaments have associated H$\alpha$ emission, implying that some regions may be in earlier stages of multiphase gas condensation. The broader, more diffuse southern filaments may trace regions where gas has only recently become thermally unstable or where cooling has been less efficient or the filaments may be dispersing, while the narrower northern filaments may represent more evolved sites of gas cooling or more efficient cooling. Additionally, projection effects and orientation relative to our line of sight could obscure the presence of narrower substructures in the southern filaments, making them appear wider.
	
	\subsubsection{M87}
	We also show filaments in M87 in Figure~\ref{fig:m87}. These filaments are co-spatial with UV and X-ray filaments \citep[e.g.,][]{werner10,anderson18,olivares25,tamhane25}. The filaments in M87 are located within the inner 3 kpc region wrapped around radio jets. The filaments are remarkably narrow with thicknesses ranging from 16 to 60 pc with an average thickness of 37 pc. Ignoring the broader filament in region 1, the average filament thickness is only $\sim$27 pc. These fine structures highlight the unprecedented resolving power of HST and suggest that the filaments are composed of narrow, magnetically confined threads, consistent with theoretical expectations for magnetic pressure support in cool cores. The significantly narrow filaments in M87, with widths of 16--25 pc, suggest that they may be denser than broader filaments. This is consistent with the higher electron densities inferred from [S \textsc{ii}] line ratios, with $\sim 600$ cm$^{-3}$ in filaments closer to the nucleus compared to $\sim 80$ cm$^{-3}$ in the outer filaments \citep{boselli19}.
	
	The optical filaments in M87 are also mostly devoid of molecular gas. The exception is the filament associated with Region 1 (see Figure~\ref{fig:m87}), which contains $\sim 4.7 \times 10^5$ M$_\odot$ of molecular gas detected by ALMA \citep{simionescu18}, likely influenced by the nearby radio jet. This region is also co-spatial with [C \textsc{ii}] emission \citep{werner13}, indicating the presence of multiphase gas and reinforcing the possibility of the presence of cold atomic gas in the filaments in NGC 5044.
	
	\begin{figure*}
		\includegraphics[width=\textwidth]{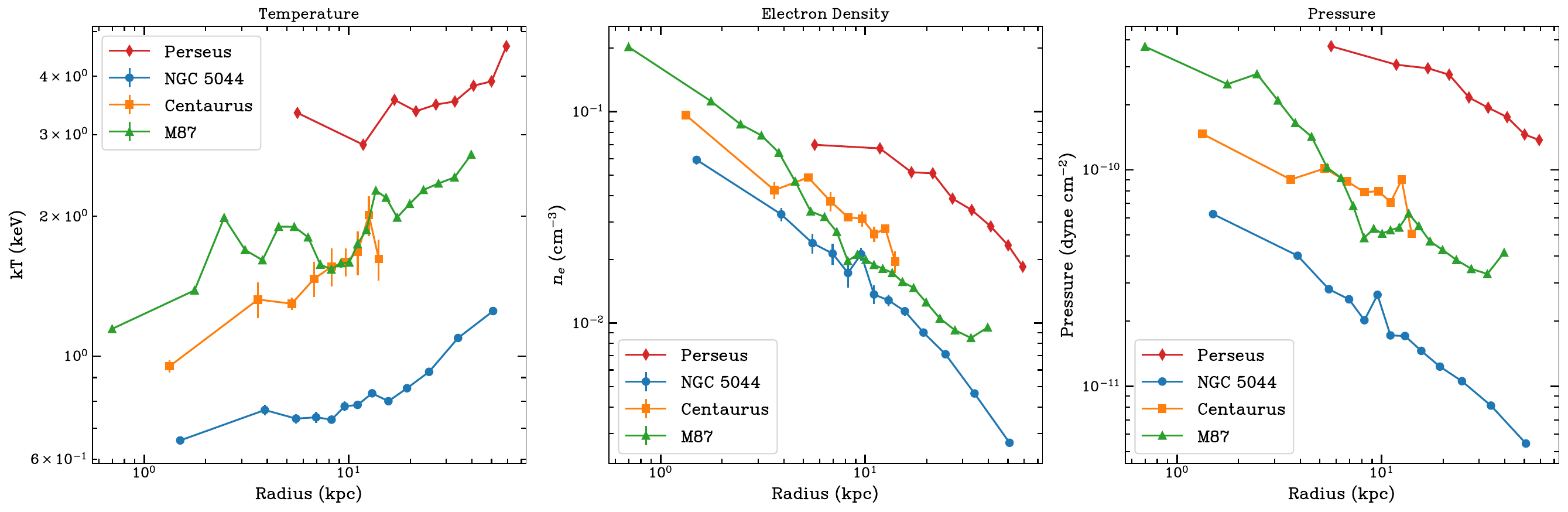}
		\caption{Comparison of deprojected temperature, electron density and pressure profiles for the four systems analysed in this paper. Perseus: \citet{sanders04}, NGC4696: \citet{babyk18}, NGC 5044 and M87: \citet{pulido18}.}
		\label{fig:prof_comp}
	\end{figure*}
	
	\subsubsection{Perseus and Centaurus}
	We re-analysed the HST data of the Centaurus cluster in a similar way as that for NGC 5044 and measured filament widths. The residual image showing the filaments is shown in Figure~\ref{fig:ngc4696}. The filaments in Centaurus extend $\sim$ 7 kpc from the centre. Their lengths vary from $\sim  2$ to 4 kpc and their widths range from 45--66 pc, comparable to value of 60 pc reported by \citet{fabian16}.
	
	For Perseus, we use the values reported in \citet{fabian08} of 70 pc width and a length of 6 pc in this analysis.
	
	Cold molecular and atomic gas has been detected in both BCGs in both the centre and in filaments traced by CO \citep{salome08,olivares19} and [C{\sc ii}] \citep{mittal11,mittal12}. However, in Centaurus, most ionised gas filaments are devoid of molecular gas. The extended molecular gas is found co-spatial with dusty filaments.

	\subsection{Filament densities}
	For NGC 5044 filaments within 1.5 kpc of the nucleus where most of the molecular gas is observed, we use the H$_2$ column density of $N_{\rm H_2} \sim 1.5 \times 10^{20}$ cm$^{-2}$ derived from ACA CO(2-1) absorption \citep{schellenberger20}. For outer filaments, we used the RMS sensitivity of ACA observations (1.8 mJy beam$^{-1}$) as an upper limit on molecular gas column density and derive the column density using
	\begin{equation}
	N(\rm{H}_2) = X_{\rm CO,gal}\,W,
	\label{eq:1}
	\end{equation}
	where $W$ is the integrated CO(1--0) line intensity in K km s$^{-1}$ and $X_{\rm CO,gal} \sim 2 \times 10^{20}$ cm$^{-2}$ (K km s$^{-1}$)$^{-1}$ is the Galactic CO-to-H$_2$ conversion factor. We assumed CO(2--1)/CO(1--0) $\sim$ 0.8 and a linewidth of 200 km s$^{-1}$ resulting in $N_{\rm H_2} < 6 \times 10^{19}$ cm$^{-2}$. $\Sigma_{\rm H_2}$ is then estimated as $2\mu_H m_pN_{\rm H_2}$, where $\mu_{H} \sim 1.4$ is the mean mass per hydrogen atom.  
	
	Alternatively, assuming that the molecular cloud in region 6 with a mass of 2.4$\times 10^6$ M$_\odot$ is associated with the filament, the estimated column density is $N_{\rm H_2} \sim 8.6 \times 10^{20}$ cm$^{-2}$ for two filaments each 2 kpc long with a width of 115 pc. These values are below the observational upper limit on the neutral atomic hydrogen column density of $N_{\rm HI} \lesssim 9.13 \times 10^{20}$ cm$^{-2}$ \citep{rajpurohit25}. We note that H$_2$ and \hi~trace different phases, so the comparison is intended to show that the total gas column does not exceed observational limits. 
	
	For comparison, the [S\,\textsc{ii}] line ratios in the filaments vary from 1.6 to 1.9 suggesting electron densities of $<30$ cm$^{-3}$ following the prescription in \citet{proxauf14}. Assuming an ionised filament with a length of 2 kpc and a thickness of 100 pc, this corresponds to a ionised gas column density upper limit of $N_{\rm H^+} < 2.16 \times 10^{20}$ cm$^{-2}$, and ionised gas mass of $<1.6 \times 10^4$ M$_\odot$ assuming a volume filling factor of 10$^{-3}$. These densities are significantly lower compared to total H column density of $N_{\rm H} \sim 2.1 \times 10^{22}$ cm$^{-2}$ estimated by ``hidden cooling flow'' model \citep{fabian23}.
	
	Similarly, in M87, $4.7 \times 10^5$ M$_\odot$ molecular gas is detected in the filament in region 1. This filament has a length of 1.4 kpc and a width of 63 pc which corresponds to $N_{\rm H_2} \sim 8.5 \times 10^{20}$ cm$^{-2}$. This filament has an electron density of 80 cm$^{-3}$ from [S \textsc{ii}] ratios \citep{boselli19}. For the assumed geometry, it would need to have a filling factor of $2.4 \times 10^{-3}$ to match the column density derived above. These values are consistent with plausible column density and filling factor ranges derived in \citet{anderson18}. For filaments without molecular gas detections, the column density upper limit derived using Equation~\ref{eq:1} is $N_{\rm H_2} < 2.4 \times 10^{20}$ cm$^{-2}$.
	
	The total molecular gas mass detected in Centaurus is $8.9 \times 10^7$ M$_\odot$ \citep{tamhane22}. This gas is distributed over a $\sim100$ arcsec$^2$ region, corresponding to an average column density of $N_{\rm H_2} \approx 10^{21}$ cm$^{-2}$. However, most of the molecular gas is co-spatial with the dust lanes, and no molecular gas is detected in the region used to extract the filament dimensions (Figure~\ref{fig:ngc4696}) at the limits of those ALMA observations. Using a molecular gas upper limit of 0.43 mJy beam$^{-1}$ and Equation~\ref{eq:1}, we derive an upper limit on the molecular gas column density in the region of the extended ionised gas filaments of $N_{\rm H_2} < 2.6 \times 10^{20}$ cm$^{-2}$.
	
	\subsection{Equipartition magnetic field strengths}
	
	\begin{figure}
		\centering
		\includegraphics[width=\linewidth]{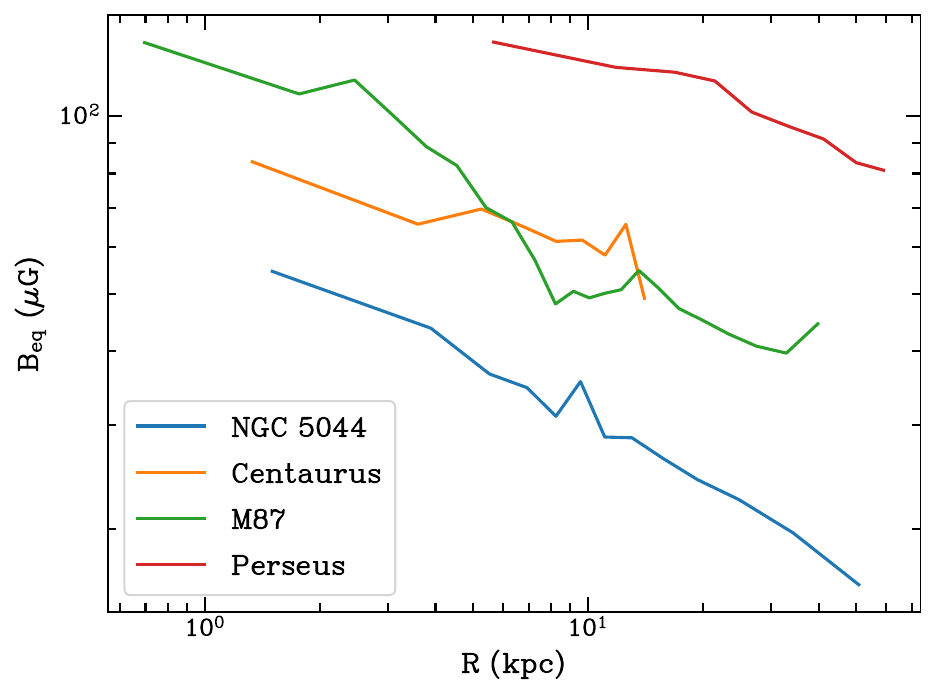}
		\caption{Equipartition magnetic field profiles of galaxies derived from their pressure profiles.}
		\label{fig:Beq}
	\end{figure}

	Several studies have shown that the multiphase filaments observed in BCGs/BGGs comprise low-entropy X-ray emitting gas ($\lesssim 1$ keV), warm molecular gas traced in the near-infrared, and ionised gas detected in the optical. These components are not in pressure equilibrium with each other and appear underpressured relative to the surrounding ICM/intra-group medium (IGM) \citep{jaffe01, oonk10, olivares25}. This pressure deficit implies the presence of additional non-thermal pressure support. Given the narrow widths and coherence of the filaments, magnetic pressure is often invoked as a leading candidate to provide this support.
	
	We first estimate the equipartition magnetic field strengths ($B_{\rm eq}$) in each system by assuming that the magnetic energy density is comparable to the total thermal energy density of the surrounding ICM, following
	\begin{equation}
	\frac{B_{\rm eq}^2}{8\pi} = (n_e + n_i)kT \approx 1.9n_e kT,
	\label{eq:2}
	\end{equation}
	where $n_e$ and $kT$ are the electron number density and temperature of the ICM/IGM. We used the temperature, electron density, and pressure profiles shown in Figure~\ref{fig:prof_comp}. We show $B_{\rm eq}$ profiles for each galaxy estimated using the above equation in Figure~\ref{fig:Beq} and show these values for the filaments in Table~\ref{tab:Bfields}. The equipartition magnetic field in NGC 5044 changes from $\sim 60 \,\mu$G in the centre to $\sim 40 \,\mu$G at 5 kpc. In M87 it varies between 180 and 100 $\mu$G in the inner 3 kpc region, and for Perseus and Centaurus it varies between 180--80 $\mu$G and 80--60 $\mu$G from centre to 50 kpc radius and 5 kpc radius, respectively.
	
	\subsection{Filament width -- pressure relationship}
	
	The differences in equipartition magnetic field strengths estimated above are consistent with theoretical expectations. In galaxy groups such as NGC~5044, the surrounding IGM is 2--5 times less dense and 3--4 times cooler than the ICM in massive clusters, resulting in ambient thermal pressures that are lower by a factor of $\sim$5--20. This reduced external confinement allows filaments to have larger transverse widths and requires less internal magnetic pressure for stability. Indeed, we observe filament widths in NGC~5044 ranging from 50--120 pc, approximately twice as wide as filaments in clusters such as Perseus, Centaurus, and M87, where widths range from 16--60 pc. This suggests that filaments in group environments are less confined and can maintain coherence with lower magnetic support, consistent with the lower equipartition magnetic fields inferred for groups. Within each system, equipartition magnetic field strengths increase by a factor of $\sim$2 in the inner regions compared to the outskirts.
	
\begin{figure}
		\centering
		\includegraphics[width=\linewidth]{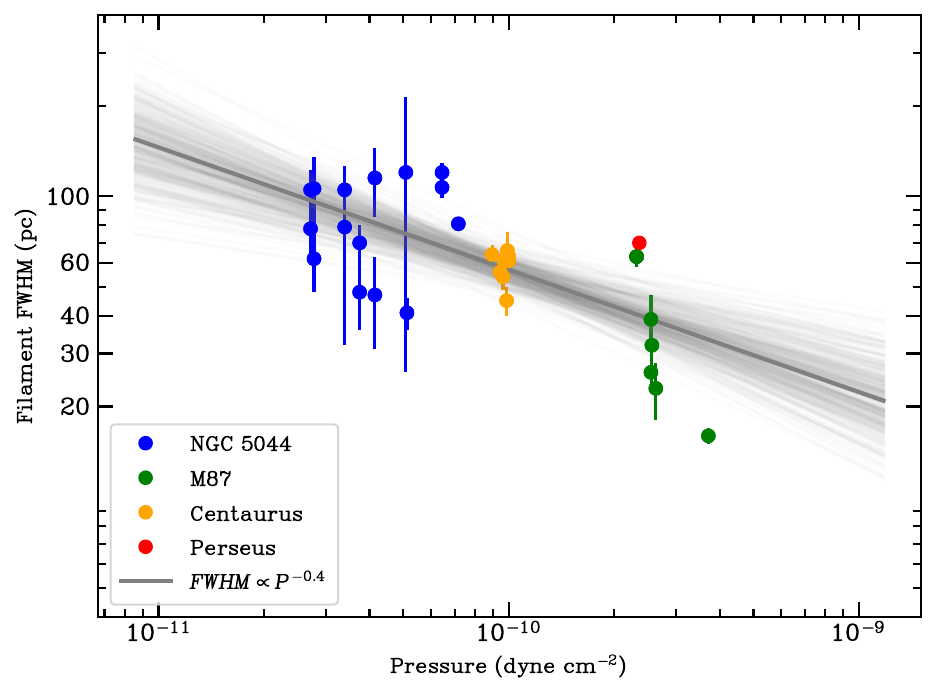}
		\caption{The relationship between filament width and pressure in galaxies in our sample. The dark grey line shows the median posterior fit, while the light grey lines show individual posterior draws.}
		\label{fig:p-width}
	\end{figure}

	To quantitatively examine this trend, we illustrate the relationship between the ambient ICM/IGM pressure ($P$) and filament width (FWHM) in Figure~\ref{fig:p-width} for NGC~5044 and for the comparison cluster sample. We modelled the relation between $\log{\rm FWHM}$ and $\log P$ as a linear function with intrinsic scatter using the Bayesian linear regression method of \citet{kelly07}, as implemented in {\sc linmix}\footnote{\url{https://github.com/jmeyers314/linmix}}. The fit accounts for measurement uncertainties in $\log{\rm FWHM}$ and intrinsic scatter. The dark grey line shows the median posterior fit, while the light grey lines show individual posterior draws, showing the full uncertainty in the slope, intercept, and intrinsic scatter. The best-fit relation is
	\begin{equation}
	\log{\rm FWHM} = (-0.40 \pm 0.09)\,\log P - (2.29 \pm 0.95),
	\end{equation}
	with an intrinsic scatter of $0.16 \pm 0.03$ dex. This result indicates that filament widths decrease with increasing ICM/IGM pressure. However, because this analysis requires deep HST narrow-band imaging that resolves extended ionised filaments at high spatial resolution, only a small number of nearby cool-core systems currently provide suitable data. The comparison sample is therefore limited to these well-studied systems. In addition, M87, in which the thinnest filaments are observed, is 2--4 times closer than the other systems, so distance and the resolution power of HST may play an important role. Broader filaments in more distant systems could in fact be composed of multiple narrower filaments that would be resolved if observed at comparable physical resolution. Projection effects may also contribute, as some of the narrowest filaments could be viewed close to edge-on. Therefore, this trend is tentative and a larger sample spanning a wider range of pressures is required to establish this relationship more robustly.
	
	\subsection{Magnetic field estimates for dynamic stability}
	We also estimated the magnetic field strengths required to stabilise filaments against gravitational and tidal forces, following the framework of \citet{fabian08}. The magnetic field strengths required for a horizontal filament supported against vertical collapse by gravity ($B_h$) and for support against tidal stretching for radially oriented filaments ($B_r$) is given by:
	\begin{equation}
	B_h \sim \sqrt{4\pi \Sigma_\perp g} \quad \text{and} \quad B_r \sim \sqrt{4\pi \Sigma_\parallel g_{\rm tidal}},
	\label{eq:3}
	\end{equation}
	where $g$ is the local gravitational acceleration, $g_{\rm tidal} = (l/R) g $ is the tidal acceleration where $R$ is the filament distance from the centre of the galaxy, $\Sigma_\perp$ is the surface density perpendicular to the filament axis and $\Sigma_\parallel = l/d \times \Sigma_\perp$ is the effective surface density along the filament, with $l$ and $d$ denoting the length and thickness of the filament, respectively \citep{fabian08}. Note that the calculation of $B_r$ assumes that filaments are falling when using $g_{\rm tidal}$. If $g$ is used instead, the values of $B_r$ will be $\sqrt{l/d}$ times larger than $B_h$, which can result in $B_r$ values larger by a factor of 10 compared to $B_h$. Since $B_r$ scales inversely with filament radius, these estimates emphasise the need for high spatial resolution observations, such as those provided by the HST, to properly constrain filament geometry and stability.
	
	These magnetic field estimates are limited by uncertainties on filament geometry and projection, local pressure, filament column density and the inner mass profiles.
	Therefore, if the estimated magnetic fields differ by less than a factor of $\sim 2$, the difference may not be physically significant given the uncertainties.
	
	In Table~\ref{tab:Bfields} we show the filament lengths, widths and the derived magnetic field estimates, where we used gravitational acceleration profiles derived in \citet{pulido18} to estimate $g$. The $B_h$ values in the inner filaments in NGC 5044 are 42 $\mu$G, whereas for outer filaments they are below 20 $\mu$G. The $B_r$ values are much larger at $\gtrsim 100$ $\mu$G for inner filaments and $\lesssim 100$ $\mu$G for outer filaments. Thus, filaments located closer to the galaxy centre, where gravitational and tidal forces are stronger, appear to require higher magnetic field strengths, potentially exceeding equipartition, to prevent disruption and maintain structural integrity. 
	
	We perform a similar analysis for M87, Perseus and Centaurus. Column density values used for these calculations are provided in Table~\ref{tab:Bfields}. We used the stellar velocity profile from \citet{oldham18} to estimate $g$ using $g \sim v^2 / R$, where $R$ is the distance of the filament from the centre in M87. For Perseus, the values of filament morphology, column density and gravitational acceleration are taken from \citet{fabian08}. For Centaurus, gravitational acceleration values are derived using the relation $g = 1.4 \times 10^{-7} (r/{\rm kpc})^{-0.68}$ cm s$^{-2}$ taken from \citet{sanders16}. All of these values are reported in Table~\ref{tab:Bfields}. The values of $B_h$ are comparable to the equipartition magnetic field strength in all galaxies. Whereas $B_r$ values exceed $B_{\rm eq}$ by a factor of 2--4 in all galaxies and lie between 150 and 250 $\mu$G.
	
	\begin{table*}
		\centering
		\begin{tabular}{ccccccccccc}
			\hline
			(1)    &  (2)  &  (3)  & (4) & (5) & (6) &   (7)     &     (8)      & (9)   &   (10)   &   (11)   \\
			Region & $d_1$ & $d_2$ & $l$ & $R$ & $g$ & $N_\perp$ & $B_{\rm eq}$ & $B_h$ & $B_{r1}$ & $B_{r2}$ \\
			& (pc) & (pc) & (kpc) & (kpc) & (cm s$^{-2}$) & (cm$^{-2}$) & ($\mu$G) & ($\mu$G) & ($\mu$G) & ($\mu$G) \\
			\hline
			\multicolumn{11}{c}{\bf NGC 5044} \\
			1 & 41 $\pm$ 5   & --           & 3.5 & 2.7 & 1.31$\times10^{-7}$ & $<6 \times 10^{19}$  & 49 & $<22$ & $<225$ & -- \\
			2 & 47 $\pm$ 16  & 115 $\pm$ 30 & 3.5 & 3.8 & 1.02$\times10^{-7}$ & $<6 \times 10^{19}$  & 44 & $<19$ & $<158$ & $<101$ \\
			3 & 48 $\pm$ 12  & 70 $\pm$ 10  & 2.7 & 4.3 & 9.44$\times10^{-8}$ & $<6 \times 10^{19}$  & 42 & $<18$ & $<110$ & $<91$ \\
			4 & 79 $\pm$ 47  & 105 $\pm$ 8  & 2.1 & 4.7 & 8.80$\times10^{-8}$ & $<6 \times 10^{19}$  & 40 & $<18$ & $<62$  & $<54$ \\
			5 & 81 $\pm$ 2   & --           & 0.8 & 0.5 & 2.04$\times10^{-7}$ & $1.5 \times 10^{20}$ & 59 & 42    & 158    & -- \\
			6 & 107 $\pm$ 8  & 120 $\pm$ 9  & 2.4 & 1.3 & 2.04$\times10^{-7}$ & $1.5 \times 10^{20}$ & 56 & 42    & 279    & 264 \\
			7 & 120 $\pm$ 94 & --           & 3.1 & 2.8 & 1.30$\times10^{-7}$ & $<6 \times 10^{19}$  & 49 & $<21$ & $<118$ & -- \\
			8 & 78 $\pm$ 19  & 105 $\pm$ 17 & 2.8 & 6   & 7.56$\times10^{-8}$ & $<6 \times 10^{19}$  & 36 & $<16$ & $<67$  & $<58$ \\
			9 & 62 $\pm$ 14  & 106 $\pm$ 29 & 2.7 & 5.7 & 7.81$\times10^{-8}$ & $<6 \times 10^{19}$  & 36 & $<17$ & $<76$  & $<58$ \\
			
			\multicolumn{11}{c}{\bf M87} \\
			1 & 63 $\pm$ 2 & 63 $\pm$ 5 & 1.4  & 2.9  & 6.98$\times10^{-8}$ & 8.5$\times10^{20}$   & 105  & 59     & 194     & 194 \\
			2 & 26 $\pm$ 1 & 39 $\pm$ 8 & 1.16 & 1.9  & 1.07$\times10^{-7}$ & $<2.4 \times10^{20}$ & 110  & $<39$  & $<203$  & $<311$ \\
			3 & 16 $\pm$ 1 & --         & 0.5  & 0.25 & 1.59$\times10^{-6}$ & $<2.4 \times10^{20}$ & 143  & $<150$ & $<1184$ & -- \\
			4 & 32 $\pm$ 8 & --         & 0.31 & 1.7  & 1.49$\times10^{-7}$ & $<2.4 \times10^{20}$ & 110  & $<46$  & $<61$   & -- \\
			5 & 23 $\pm$ 5 & --         & 0.65 & 2.6  & 7.79$\times10^{-8}$ & $<2.4 \times10^{20}$ & 112  & $<33$  & $<88$   & -- \\
			
			\multicolumn{11}{c}{\bf Centaurus} \\
			1  & 61 $\pm$ 4  & --         & 1.52 & 5.8 & 4.24$\times10^{-8}$ & $<2.6 \times10^{20}$ & 68 & $<25$ & $<65$  & --    \\
			2  & 45 $\pm$ 5  & --         & 1.3  & 4.8 & 4.80$\times10^{-8}$ & $<2.6 \times10^{20}$ & 69 & $<27$ & $<76$  & --    \\
			3  & 64 $\pm$ 5  & --         & 2.13 & 6.7 & 3.86$\times10^{-8}$ & $<2.6 \times10^{20}$ & 65 & $<24$ & $<79$  & --    \\
			4  & 61 $\pm$ 4  & 63 $\pm$ 2 & 1.32 & 5.0 & 4.67$\times10^{-8}$ & $<2.6 \times10^{20}$ & 69 & $<27$ & $<64$  & $<63$ \\
			5  & 54 $\pm$ 5  & --         & 2.25 & 3.4 & 6.14$\times10^{-8}$ & $<2.6 \times10^{20}$ & 68 & $<31$ & $<162$ & --    \\
			6  & 56 $\pm$ 2  & --         & 2.27 & 4.2 & 5.28$\times10^{-8}$ & $<2.6 \times10^{20}$ & 67 & $<28$ & $<133$ & --    \\
			7  & 66 $\pm$ 10 & --         & 2.25 & 4.9 & 4.76$\times10^{-8}$ & $<2.6 \times10^{20}$ & 69 & $<27$ & $<107$ & --    \\
			
			\multicolumn{11}{c}{\bf Perseus} \\
			1  & 70 & -- & 6 & 25 & 6.35$\times10^{-8}$ & 4$\times10^{20}$ & 106 & 39 & 175 & -- \\
			
			\hline
		\end{tabular}
		\caption{The table reports the properties of the filaments. Column (1) lists the regions used to extract the surface brightness profiles shown in Figures~\ref{fig:N5044} and \ref{fig:m87}. Columns (2)--(3) give the filament FWHM within each region. Column (4) gives the filament length, (5) the projected distance from the galaxy centre, and (6) the gravitational acceleration at that distance. Column (7) lists the perpendicular filament column density used to estimate magnetic fields. Columns (8)--(9) report the equipartition and horizontal magnetic field components, while Columns (10)--(11) give the radial magnetic field components for filaments in each region estimated using Equation~\ref{eq:3}.
		}
		\label{tab:Bfields}
	\end{table*}

	\subsection{Dust}
	\label{sec:dust}
	
	\begin{figure}
		\centering
		\includegraphics[width=\linewidth]{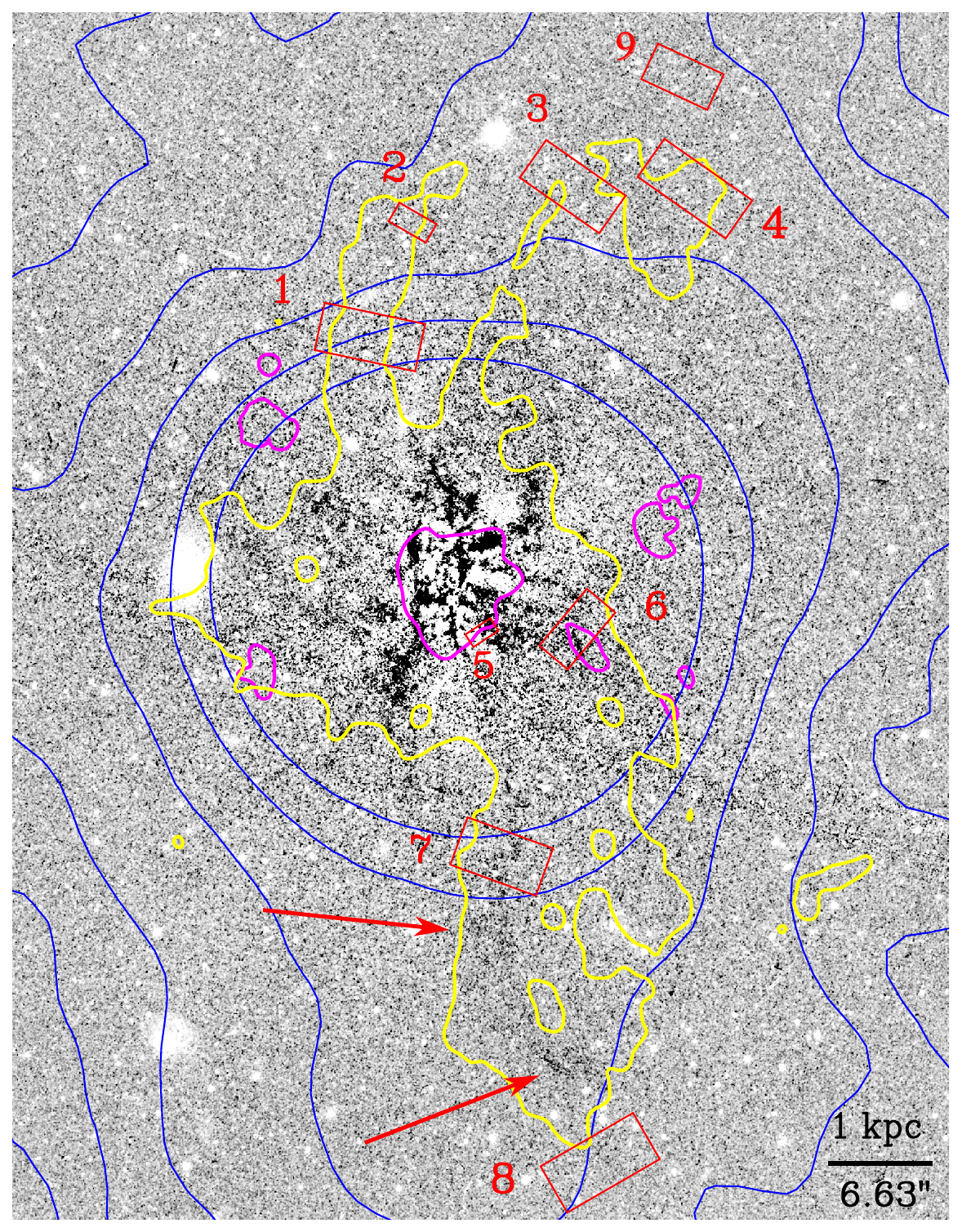}
		\caption{HST F814W residual image showing dust in absorption (darker regions have more negative pixel values, or more dust extinction). The blue contours trace the radio jet emission with GMRT and magenta contours are the CO(2-1) flux density at 0.17 Jy km s$^{-1}$ in moment 0 maps. The yellow contour shows the 2.5$\sigma$ H$\alpha$ emission. The red arrows indicate the faint dust cospatial with the southern filament. Regions used for extracting H$\alpha$ filament widths are shown as red boxes as in Figure~\ref{fig:N5044}. Regions 9 and 8 are outside the MUSE field of view.}
		\label{fig:n5044_dust}
	\end{figure}
	
	Figure~\ref{fig:n5044_dust} shows the F814W residual map, highlighting dust extinction features in NGC 5044. Several dusty filaments are visible within the central 1--2 kpc, including structures very close to the nucleus. Dust extinction is also visible along the southern filament indicated by red arrows, in the direction of radio lobes, whereas no dust is detected in the northern filaments. This asymmetry may be explained by filament orientation, with the southern filament located on the near side along the line of sight and the northern filaments lying on the far side. However, dust extinction is strongest in the inner $\sim 1$ kpc region as shown in the right panel of Figure~\ref{fig:N5044_cen}. We estimated the extinction in the F814W band in the strongest features in the centre using $A_{814} = 2.5 \,{\rm Log((F814_{mod} - abs(F814_{res}))}$ / ${\rm F814_{mod})}$,
	where $F814_{\rm mod}$ and $F814_{\rm res}$ are the fluxes extracted from the model galaxy continuum image and the residual image, respectively, within the dust absorption regions. The visual extinction $A_V$ was then derived using the \citet{cardelli89} extinction law with $R_V = 3.1$, yielding an average $A_V = 0.07 \pm 0.03$.
	
	To further constrain the dust properties, we used the $Spitzer$ 70 and 160 $\mu$m fluxes reported by \citet{temi09} and fit a modified blackbody spectrum with the {\sc mbb\_emcee}\footnote{\url{https://github.com/aconley/mbb_emcee}} code, which employs an affine-invariant Markov Chain Monte Carlo (MCMC) method. We restricted the allowed parameter ranges to $T_{\rm dust} = 10$--$40$~K and dust emissivity index $\beta = 1$--$3$. For NGC~5044, this yields a dust mass of $M_{\rm dust} = 3.2^{+18.4}_{-2} \times 10^5$~M$_\odot$, an 8--1000~$\mu$m infrared luminosity of $L_{\rm IR} = 2.2 \pm 0.5 \times 10^{42}$~erg~s$^{-1}$, a dust temperature $T_{\rm dust} = 32.2^{+7.1}_{-12.3}$~K, and $\beta = 1.7 \pm 0.4$. This corresponds to a dust-to-gas mass ratio of $0.003^{+0.018}_{-0.002}$, assuming a total molecular gas mass of $10^8$~M$_\odot$ \citep{schellenberger20}.  
	
	Performing a similar analysis for NGC~4696 using FIR fluxes from $Spitzer$ (24, 70 and 160 $\mu$m) and $Herschel$ SPIRE (250, 350 and 500 $\mu$m) retrieved from NED\footnote{\url{https://ned.ipac.caltech.edu/}}, we find $M_{\rm dust} = 4.1^{+2.2}_{-1.1} \times 10^5$~M$_\odot$, $L_{\rm IR} = 2.5^{+0.2}_{-0.3} \times 10^{42}$~erg~s$^{-1}$, $T_{\rm dust} = 28.7^{+4.8}_{-4.1}$~K, and $\beta = 2.0 \pm 0.2$, corresponding to a dust-to-gas mass ratio of $0.005^{+0.002}_{-0.001}$ for a total molecular gas mass of $\sim9 \times 10^7$ M$_\odot$ \citep{tamhane22}. However, \citet{mittal11} estimated cold dust ($T_{\rm dust} = 19$ K) mass of $1.6 \times 10^6$ M$_\odot$ using two-temperature fit to the IR spectrum. Assuming this dust mass, the dust-to-gas ratio would increase to 0.018.
	
	For comparison, in the Perseus cluster, the dust mass is $\sim 6 \times 10^7$~M$_\odot$ (with $T_{\rm dust} \sim 20$~K and $\beta = 1.3$) and the total H {\sc i}+H$_2$ mass is $\sim 10^{10}$~M$_\odot$, giving a dust-to-gas mass ratio of $\sim 0.006$ \citep{salome11}.
	
	These dust-to-gas mass ratios are lower than the typical Milky Way value of $\sim 0.01$ \citep{draine11}, but comparable to dust-to-gas ratios in other similar central galaxies in groups and clusters \citep[e.g.,][]{edge01}.

	\section{NUCLEAR REGION OF NGC~5044}
	\label{sec:nuc_reg}
	
	\begin{figure*}
		\centering
		\includegraphics[width=\linewidth]{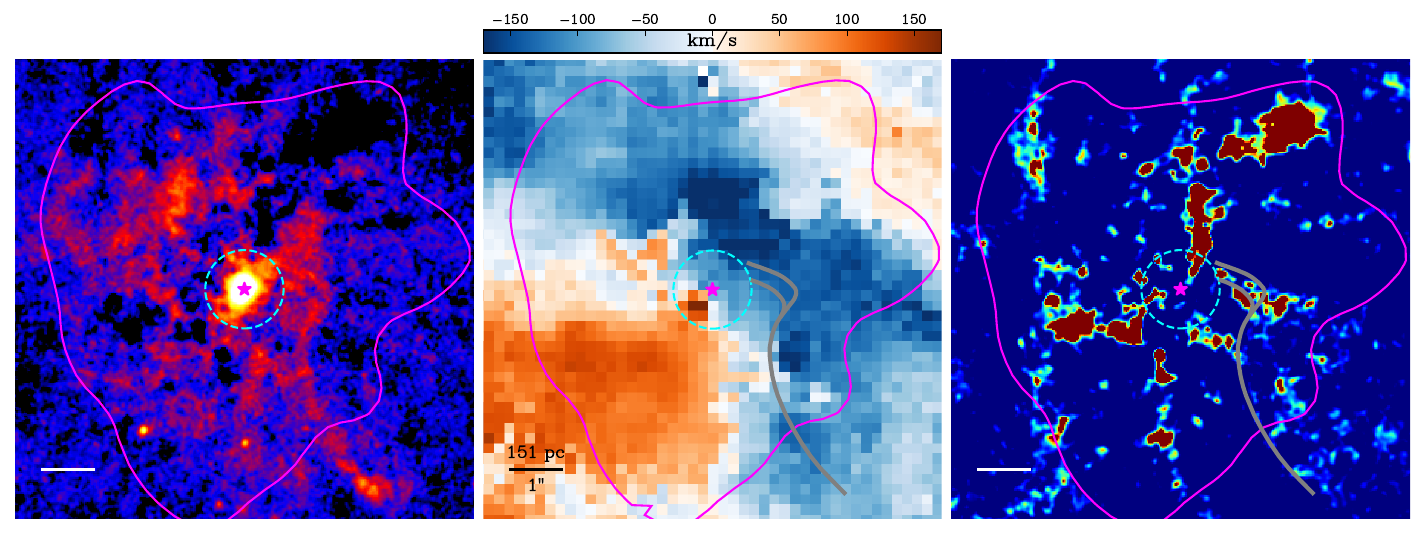}
		\vspace{-0.7cm}
		\caption{The central region of NGC 5044 is shown in the net H$\alpha$+N[\textsc{ii}] (HST F665N residual), MUSE H$\alpha$ velocity map and F814W residual map (see section~\ref{sec:da_hst} for details) to show the ionised gas morphology, velocity and dust distribution in the left, middle and right panels, respectively. Bright ionised gas core and an extended filament connecting within the ionised gas core is visible in the left panel.
			The dashed cyan circles show the Bondi radius of NGC 5044 (116 pc). The magenta stars show the location of the AGN. The ALMA CO(2-1) emission contours tracing the cold molecular gas at 0.17 Jy km s$^{-1}$ in moment 0 maps are shown in magenta. The grey curves in the middle and right panels show the $\sim$800 pc long inner filament connecting with the ionised gas core, identified in the H$\alpha$+N[\textsc{ii}] image in the left panel.
		}
		\label{fig:N5044_cen}
	\end{figure*}
	
	Figure~\ref{fig:N5044_cen} shows the nuclear region of NGC 5044 in HST F665N, MUSE H$\alpha$ velocity map and F814W residual maps. The high spatial resolution of the HST resolves the Bondi radius of $\sim 116$ pc. To estimate the Bondi radius, we adopt $R_{\rm B} = 2GM_{\rm BH}/c_s^2$ ($\gamma \sim 5/3$), where $G$ is the gravitational constant, $M_{\rm BH}$ is the mass of the central supermassive black hole, and $c_s$ is the sound speed of the ambient gas. Adopting $R_{\rm B} = GM_{\rm BH}/c_s^2$ would halve the radius. \citet{diniz17} estimated $M_{\rm BH} \sim 1.8 \times 10^9$ M$_\odot$ from integral field spectroscopy of the nuclear region using the updated M--$\sigma$ relation. Assuming the gas temperature in the inner 1 kpc is 0.5 keV, the sound speed is $c_s = \sqrt{\gamma k T/\mu m_p} \approx 365$ km s$^{-1}$, yielding a Bondi radius of 116 pc. The corresponding Bondi accretion rate given by $\dot{\rm M} = \pi \rho G^2 M^2/ c_s^3$ is $\sim 0.03$ M$_\odot$ yr$^{-1}$ under the assumption of spherical symmetry and negligible angular momentum, comparable to the accretion rate of $\sim 0.01$ M$_\odot$ yr$^{-1}$ derived by \citet{schellenberger24} from spectral energy distribution (SED) modelling of the nuclear radio source using a JP model for the pc-scale jet and an Advection-Dominated Accretion Flow (ADAF) model for the accretion disk. We also estimated the SMBH sphere of influence as $G M_{\rm BH}/\sigma^2 \sim 153$ pc for a $\sigma \sim 225$ km s$^{-1}$.
	
	We detect an ionised gas core within the Bondi radius, centred on the SMBH. Its radial intensity profile peaks at the nucleus and declines roughly as $r^{-1.07}$ out to $\sim$100 pc as shown in Figure~\ref{fig:f665_sbprof} in Appendix~\ref{f665n_profile}. The lower-resolution MUSE H$\alpha$ velocity map shows hints of rotation around the SMBH, which could indicate a circumnuclear disk similar to those observed in other systems. The CO line velocity field also looks like rotation, consistent with the velocity field of ionised gas \citep{schellenberger20}. Although radial inflows or outflows cannot be ruled out. This compact, dense ionised core appears to be connected to more extended structures, as revealed by filamentary features extending hundreds of parsecs from the nucleus. We identify a $\sim$800 pc long filament connected to the inner core to the northwest, similar to the swirling filaments observed in Centaurus and M87 \citep{fabian16,ford94}. We also detected $\sim$40--80 pc wide dust lanes in projection within the core (right panel of Figure~\ref{fig:N5044_cen}). The narrow widths of these structures suggest they are magnetically confined, channeling gas along field lines toward the nucleus and potentially feeding the central black hole. Similar filamentary inflows have been observed in the Perseus cluster, where molecular gas filaments are thought to supply the circumnuclear disk (within $\sim$100 pc) surrounding the SMBH, which may in turn drive accretion onto the black hole \citep{oosterloo24}.
	
	Molecular gas clouds have been detected in absorption against the continuum, with velocities of $\sim$260 km s$^{-1}$ and masses of 6--7$\times10^3$ M$_\odot$ within $\sim$20 pc of the SMBH, well inside the Bondi radius, suggesting potential accretion \citep{schellenberger20}. However, an accretion efficiency of less than 50\% is required to reconcile the observed energy output (from X-ray cavities, radio emission, and bolometric AGN luminosity) with this inflow rate, implying that only a small fraction of the circumnuclear disk mass reaches the black hole. Similar results have been found in M87, where analyses of ionised gas filaments around its SMBH reveal that not all filaments pass close to the black hole \citep{osorno23}. Neutral atomic gas has also been detected in absorption against the continuum in NGC 5044, with velocities correlated with those of the CO(2--1) absorption lines, although with broader dispersions \citep{rajpurohit25}. This indicates the presence of neutral atomic phase within the circumnuclear environments, kinematically linked to the molecular gas, highlighting the multiphase nature of the accretion flow onto the SMBH.
	
	The H$\alpha$ flux within a 116 pc (0.77$''$) aperture centred on the nucleus in the MUSE maps is $4.2\times 10^{-15}$ erg s$^{-1}$ cm$^{-2}$, after subtracting the contribution from an annular background aperture with $r_{\rm in} = 0.8''$ and $r_{\rm out} = 2.3''$. This corresponds to an H$\alpha$ luminosity of $4.9 \times 10^{38}$ erg s$^{-1}$. The luminosity can also be expressed as \citep{osterbrock06}
	\begin{equation}
	L_{\rm H\alpha} = \gamma n_e^2 f V,
	\end{equation}
	where $\gamma \sim 3 \times 10^{-25}$ erg cm$^3$ s$^{-1}$ is the case B recombination emissivity coefficient at 12,000 K, $n_e$ is the electron density, $f$ is the volume filling factor, and $V$ is the emitting volume. Using the [S\,\textsc{ii}] line ratio of $1.37 \pm 0.04$,
	we estimate an average electron density of $\lesssim 70$ cm$^{-3}$ within the Bondi-radius aperture, following the updated calibration of \citep{proxauf14}. Assuming spherical geometry and this density, the filling factor is $f \gtrsim 3 \times 10^{-3}$, which implies an ionised gas mass of $<2.8 \times 10^4$ M$_\odot$ within the Bondi radius. The molecular gas mass estimated within the same aperture from ALMA CO(2-1) spectral cube gives $M_{\rm CO} \sim 2.8 \times 10^6$ M$_\odot$. This is the total gas mass along the line of sight and it is not clear how much of the gas lies within the Bondi radius. Given that the H$\alpha$+[N \textsc{ii}] surface brightness scales approximately as $I \propto r^{-1.07}$ within this region, the corresponding electron density profile follows $n_e \propto r^{-1.07}$, assuming constant emissivity coefficient, volume filling factor and H$\alpha$/[N \textsc{ii}] ratio throughout the Bondi radius. This slope is close to the $n_e \propto r^{-0.8}$ inferred by \citet{david09} from Chandra observations of hot gas in the central region. The density profile is shallower than the $r^{-1.5}$ scaling expected from analytic Bondi solution, but comparable to CCA \citep{gaspari13}, more recent gravitational-MHD (GRMHD) simulations \citep[e.g.,][]{guo23,xu23,cho25} and $r^{-1}$ slope estimated in M87 and M84 from X-ray observations of hot gas within their Bondi radii \citep{russell15,bambic23}.

	\section{Discussion}
	\subsection{Comparison with simulations and radio measurements}
	It is useful to compare these inferred values with direct observational constraints from radio studies. The inferred equipartition magnetic field strengths are higher than the $\sim$10--25\,$\mu$G values derived in several radio-loud cool-core clusters from Faraday rotation measure (RM) \citep[e.g.,][]{taylor01,taylor07,allen01,feretti99}. Whereas in galaxy groups, the RM-derived magnetic field strengths are of the order of a few $\mu$G with a mean of 5 $\mu$G \citep{anderson24}. More recently, \citet{rajpurohit25} estimated a magnetic field strength of $\sim$2\,$\mu$G in the radio lobes of NGC 5044 using energy equipartition arguments, which is an order of magnitude lower than the filament $B_{\rm eq}$. However, Faraday RMs are sensitive to the integrated line-of-sight magnetic field component weighted by the electron density. Since angular resolution of these observations are often of the order of a few arcseconds to arcminutes, corresponding to kiloparsec or larger scales, the magnetic field strength derived from RM tends to reflect larger-scale, volume-averaged fields, not necessarily the strongest fields within the thin filaments which may not be aligned along the line-of-sight and are not volume filling. Polarised radio observations with high sensitivity and high-resolution with beam sizes matching the widths of filaments may be required to probe the magnetic field strengths in these structures.
		
	On the other hand, recent MHD and GRMHD simulations predict field strengths of $\sim 10^2~\mu$G in cold ($T \lesssim 10^4$ K) filaments, with field lines preferentially aligned along the filament major axis in magnetically supported structures \citep[e.g.,][]{fournier24,guo24}. These values are consistent with the radial magnetic field strengths ($B_r$) inferred for the filaments in our sample. If in reality such high levels of magnetic fields exist in filaments, they can be explained by flux freezing and magnetic draping. For example, if the filaments condense from uplifted low-entropy gas, as appears to occur in these systems, flux freezing during cooling naturally amplifies the magnetic field in the denser cold phase. For example, a low-entropy ($T \lesssim 0.5$ keV) filament of width $w_{\rm{hot}} = 200$~pc and $B_{r,\,\rm{hot}} \approx 10~\mu$G compressed into a cold filament of width $w_{\rm{cold}} = 60$~pc would, under flux freezing, reach
	\begin{equation}
	B_{r,\,\rm{cold}} \approx B_{r,\,\rm{hot}} \left( \frac{w_{\rm{hot}}}{w_{\rm{cold}}} \right)^2 \approx 111~\mu\rm{G},
	\end{equation}
	comparable to the values in Table~\ref{tab:Bfields}. This assumes ideal MHD and conservation of magnetic flux perpendicular to the filament length. This simple scaling illustrates that even moderate geometric compression of an initially modest magnetic field can produce the $\sim 10^2~\mu$G fields inferred in the cold phase, providing sufficient magnetic pressure to contribute significantly to filament stability and confinement.
	
\begin{figure}
		\includegraphics[width=0.48\textwidth]{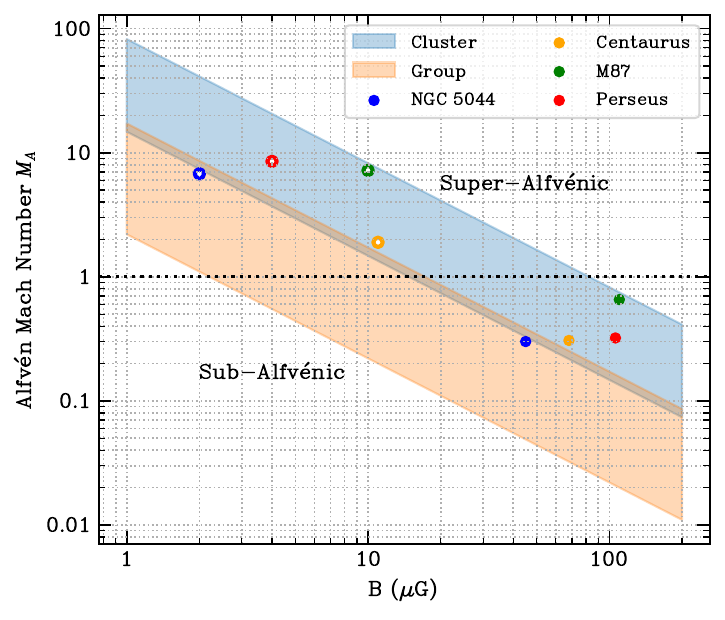}
		\caption{Alfv\'en Mach number ($M_A$) as a function of magnetic field strength ($B$) in the IGM/ICM. Blue and orange shaded regions represent typical cluster and group environments, respectively. The $M_A$ ranges are derived assuming electron densities of $n_e$ = 0.02--0.1 cm$^{-3}$ for clusters and $n_e$ = 0.001--0.02 cm$^{-3}$ for groups, and turbulent velocity dispersions of $v_{\rm turb}$ = 120--300 km s$^{-1}$ for clusters and $v_{\rm turb}$ = 80--140 km s$^{-1}$ for groups. Alfv\'en Mach numbers estimated for galaxies in our sample are shown as solid symbols when computed using $B_{\rm eq}$, and as open symbols when estimated using large-scale magnetic field strengths inferred from radio observations. The black horizontal dotted line at $M_A = 1$ shows the separation between sub-Alfv\'enic and super-Alfv\'enic turbulence regimes. }
		\label{fig:B-turb}
	\end{figure}

	If ambipolar diffusion is sufficiently slow, the molecular, atomic, and ionised components can be linked to each other via magnetic fields \citep[e.g.,][]{fabian08}. In that case, the measured internal velocity dispersions of $\sim$ 100 km s$^{-1}$ could be consistent with Alfv\'enic turbulence, where roughly half of the internal pressure is kinetic and half magnetic, providing modest internal support within the filament. Recent X-ray spectroscopic observations with \textit{XRISM} have measured line-of-sight turbulent velocities of $v_{\rm turb} \sim 150$ km s$^{-1}$ in the ICM \citep{hitomi16,xrism_perseus,xrism_centaurus,xrism_m87}, which corresponds to 3D turbulent velocity dispersion of $\sigma_{\rm turb} \sim 260$ km s$^{-1}$. For typical cluster core densities of $n_e \sim 0.1$ cm$^{-3}$ (corresponding to $\rho \sim \mu_e m_p n_e$, where $ \mu_e \approx 1.2)$ and magnetic field strengths of $B \sim 10~\mu$G, the corresponding Alfv\'en speed ($v_A$) is
	\begin{equation}
	v_A = \frac{B}{\sqrt{4 \pi \rho}}~\sim~63~\rm{km~s^{-1}}
	\end{equation}
	yielding an Alfv\'enic Mach number of
	\begin{equation}
	M_A = \frac{\sigma_{\rm turb}}{v_A}~\sim~4.
	\end{equation}
	This places ICM turbulence in the super-Alfv\'enic regime. For lower ambient magnetic field strengths or higher densities, the flow remains super-Alfv\'enic. In group environments, where densities are 10 times lower and 1D turbulent velocities are lower \citep[$v_{\rm turb} \sim 110$ km s$^{-1}$;][]{sanders11,ogorzalek17}, the Alfv\'enic Mach number is $\sim 2$ for $B \sim 5 \,\mu$G. We illustrate this dependence in Figure~\ref{fig:B-turb}, which shows how $M_A$ varies with magnetic field strength for both group and cluster like hot gas densities and turbulent velocity dispersions. At $B \lesssim 3~\mu$G, turbulence is clearly super-Alfv\'enic even in less dense environments, implying that turbulent motions can distort magnetic structures unless the field is significantly amplified.
	
	These conditions have important implications for the structure and survival of cold filaments. In the trans- or super-Alfv\'enic regime, turbulent magnetic fluctuations are no longer negligible compared to the ordered field, potentially disrupting the stability of magnetically supported filaments. However, in the wakes of AGN-inflated cavities or around the rims of cavities where field lines are stretched and draped around the bubble surface, the local magnetic field can be amplified by factors of 2--3 \citep[e.g.,][]{lyutikov06,dursi08}. If the ambient $M_A$ is near unity (as shown in Figure~\ref{fig:B-turb}), such draping can make the local flow sub-Alfv\'enic, enhancing magnetic tension and insulating cold gas against conduction and shear. In such a regime, turbulent magnetic fluctuations are energetically subdominant compared to the large-scale field, consistent with the coherent, largely radial magnetic fields inferred in cold filaments \citep[e.g.,][]{beattie22}. Therefore, flux freezing and compression along these coherent fields can plausibly amplify the filament magnetic field to the $\sim 10^2~\mu$G levels, while turbulence plays only a minor role in disrupting the field geometry. Conversely, in more weakly magnetised, high-density or high turbulence environments, where the ambient $M_A \gg 1$, even draping may be insufficient to stabilise the filaments, leading to rapid disruption unless other processes (e.g., rapid cooling or clumping) intervene. This framework naturally explains the prevalence of long, coherent filaments in cavity wakes, while cold gas in other environments may be short-lived or appear fragmented.
	
	We note that existing RM observations generally indicate smaller large-scale magnetic fields. However, Very Long Baseline Array (VLBA) observations of the Perseus cluster reported a line-of-sight field $B_\parallel \sim 4~\mu$G, assuming the RM arises from ionised gas filaments with a path length of 10 pc \citep{taylor06}. If the path length is shorter, of order 1 pc (e.g. if the ionised gas has a small volume filling factor), the inferred $B_\parallel$ increases to $\sim 50~\mu$G, and depending on the filament orientation, the radial magnetic field could reach $\sim 100~\mu$G. Given the uncertainties in our radial field strength estimates (e.g., pressure profiles, projection effects, column densities), differences of less than a factor of $\sim$2 may not be physically significant, and recent MHD simulations predict median field strengths of $80-100$ $\mu$G in cold clumps within filaments \citep[e.g.,][]{fournier24}. Thus, the observed values may be broadly consistent if path lengths are short, though field strengths significantly above $\sim$100 $\mu$G remain difficult to reconcile with observations.

	\subsection{Non-thermal pressure support in filaments}
	
	Other non-thermal processes such as turbulence and cosmic rays can also provide the internal pressure support for filaments. Following \citet{olivares25}, the turbulent pressure can be approximated as $P_{\rm turb} = \frac{1}{3}\rho \sigma_v^2$, where $\sigma_v$ is the 3D turbulent velocity dispersion and $\rho$ is the gas density. For typical cluster filament electron densities of $n_e \sim 60$ cm$^{-3}$, volume filling factor of 0.005 and turbulent velocities of $\sim 100$ km s$^{-1}$ in filaments \citep[e.g.,][]{olivares22}, the turbulent pressure is $P_{\rm turb} \sim 2.3 \times 10^{-11}~\mathrm{dyne~cm^{-2}}$, which would reduce the required magnetic field strength by $\sim 24~\mu$G. In group environments with a density of 10 cm$^{-3}$ and similar covering fraction, $P_{\rm turb} \sim 1.2 \times 10^{-11}$ dyne cm$^{-2}$, corresponding to $\sim 17~\mu$G, implying turbulence can provide $\sim 25 - 30\%$ non-thermal support in groups and clusters. Cosmic rays may also contribute additional pressure support \citep[e.g.,][]{beckmann22}. Thus, while magnetic fields can in principle provide both pressure support and structural cohesion, the very large field strengths are difficult to reconcile with the relatively modest Faraday rotation measures and faint radio emission associated with most filaments. Therefore, turbulence and cosmic rays can provide complementary or even dominant non-thermal support.
	
	At the same time, purely hydrodynamic simulations demonstrate that the formation of multiphase gas does not require magnetic fields. For example, \citet{li14,qiu19} showed that thermal instability triggered by uplift can lead to condensation of cold gas from the hot ICM, while CCA simulations predict that turbulence alone can compress gas at the interfaces of eddies, naturally forming extended filamentary structures \citep{gaspari17}. These hydro simulations tend to produce clumpier, more sheet-like cold structures, while MHD simulations \citep[e.g.,][]{ehlert23,fournier24} show that magnetic fields can collimate and stabilise the filaments, yielding the long, coherent morphologies observed. Thus, condensation-driven models and magnetic stabilization are complementary, where turbulence and uplift promote the formation of cold gas, while magnetic fields determine its longevity and morphology which may not require such strong magnetic fields.

	\subsection{Star formation and Magnetic field strengths}
	
	To assess whether the filaments are gravitationally unstable, we compare their surface density to the magnetic critical value. The critical surface density for gravitational instability is given by \citet{mckee93} as
	\begin{equation}
	\Sigma_c = \frac{B}{2 \pi \sqrt{G}} = 0.062 \left( \frac{B}{100 \, \mu{\rm G}} \right) \, {\rm g \, cm^{-2}},
	\end{equation}
	where $G$ is the gravitational constant. Using $\Sigma = \mu m_p N$ with $\mu \sim 1$, this corresponds to column densities of approximately $0.4$--$1.5 \times 10^{22}$ cm$^{-2}$ in NGC~5044 and $1$--$3.5 \times 10^{22}$ cm$^{-2}$ in galaxy clusters for $B_{\rm eq}$. These values exceed the observed column densities in these systems by almost two orders of magnitude suggesting that the filaments are gravitationally stable. However, some dense clumps of molecular gas can reach densities of 10$^{21}$--10$^{22}$ cm$^{-2}$ since these column densities are required to excite CO molecules, comparable to the critical density, yet star formation in filaments is quenched.
	
	The right panel of Figure~\ref{fig:ngc5044_rgb} shows that no blue star clusters are detected, indicating an absence of recent star formation. The F300X--F814W colour--magnitude diagram for point sources detected on F300X image with 0.5$''$ diameter aperture shows no evidence of blue sources. A comparison of the 3$\sigma$ magnitude limits in F300X (22.83 mag) and F814W (22.52 mag) with predictions from Starburst99 models \citep{leitherer99} places an upper limit on the recent ($\lesssim 10^7$ yr) star formation rate (SFR) of $<10^{-3}$ M$_\odot$ yr$^{-1}$. A similar lack of star formation is observed in the filaments of M87 \citep{tamhane25} and the Centaurus cluster \citep[e.g.,][]{fabian24}. By contrast, star formation has been detected in some of the filaments in Perseus \citep{canning14}, showing that under certain conditions stars can form in these structures. \citet{fabian08} argued that collapse can occur if the molecular mass is concentrated in a very small thickness ($\sim 0.1$ pc). Alternatively, jets can directly compress the filaments. Indeed, in most BCGs with filamentary FUV continuum emission, the star formation appears to be associated with the radio jets \citep{crockett12,canning13,tremblay15,tamhane23}. Even in Perseus, there are indications that past jet orientations may have intersected the star-forming filaments \citep{canning10}. Simulations also predict that in the presence of jets, star formation efficiency is enhanced \citep[e.g.,][]{ehlert23}. Overall, these findings are consistent with the broader picture of suppressed star formation in BCGs, despite their large molecular gas reservoirs, much of which resides in filamentary structures \citep{tamhane22}, perhaps due to enhanced magnetic fields stabilizing the filaments, turbulence and the lack of external pressure.
	
	\subsection{Multiphase structure and molecular gas survival in filaments}
	\label{sec:fil_structure}
	Filaments are unlikely to be monolithic structures but instead may consist of a mist of small, dense cloudlets embedded within a more diffuse phase. Simulations suggest that thermal instabilities fragment cooling gas into sub-pc to tens-of-pc sized clumps with very low volume filling factors \citep[e.g.,][]{mccourt18,fournier24}. Such a multiphase internal structure naturally explains the mismatch between density estimates from different tracers. Absorption against compact continuum sources is sensitive to rare, high-density clumps, while emission measurements average over a much larger area and preferentially probe the more diffuse component. This picture is supported by absorption measurements of \citet{rose23}, which reveal significantly higher column densities than inferred from emission, indicating that molecular gas is confined to small, dense structures. \citet{schellenberger20} similarly report $\sim 10$ times higher column densities for absorbing clouds in NGC 5044 when the cloud sizes and covering fractions are taken into account. The clumpiness of the medium therefore provides a consistent picture in which most of the cold mass may reside in small, dense pockets coherently held together by magnetic fields while the bulk of the filamentary volume is filled by lower-density atomic and ionised gas.
	
	\subsubsection{The role of ambient pressure}
	The absence of strong CO in the NGC 5044 filaments can be explained in terms of pressure regulation of cold gas formation and the fragility of H {\sc i} in feedback-dominated atmospheres. Recent studies show that molecular content correlates strongly with pressure. \citet{babyk23} find that the molecular gas mass scales steeply with hydrostatic pressure ($M_{\rm mol} \propto P^{1.9}$) and that the H$_2$/H {\sc i} ratio rises with pressure, consistent with rapid conversion of H {\sc i} to H$_2$ in dense, high-pressure environments. In this framework, group filaments outside the central 1--2 kpc region may be unable to accumulate the H {\sc i} reservoirs needed to form and shield large molecular clouds, leading instead to predominantly ionised or atomic gas with only sporadic molecular clumps, whereas the higher external pressure in clusters can stabilise and rapidly grow extended molecular cloud formation.
	
	AGN feedback further limits neutral gas survival. Simulations show that AGN feedback efficiently destroys neutral hydrogen, reducing the H {\sc i} mass in halos by $\sim$50\% \citep{vn16}, making it more difficult to reach the column densities required for effective shielding against dissociation in groups. For comparison, in galactic disks, $N_{\rm H\,I}\sim10^{21}$ cm$^{-2}$ is typically required to shield H$_2$ \citep{krumholz09}. Dense molecular clumps that do form may dynamically decouple from the diffuse filamentary gas through turbulent mixing or ambipolar diffusion. In groups, such clumps are likely transient and may move ballistically, consistent with the weak spatial correlation between molecular gas and extended ionised filaments observed outside the central regions of NGC 5044. In clusters, by contrast, higher external pressures allow molecular clumps to remain closer to pressure balance, enabling longer survival and closer association with ionised filaments. Observations of massive ellipticals further support this scenario, where non-central galaxies host only compact cold gas reservoirs, while central galaxies with higher atmospheric pressures and lower cooling times host extended multiphase halos \citep{temi22}. Moreover, recent MHD simulations also find that in low-pressure halos, cooling is centrally concentrated, but in high pressure systems with stronger magnetic fields, cold gas is more extended \citep{grete25,prasad26}.

	\subsubsection{The role of dust and local conditions}
	While pressure is an important regulator, it is not sufficient on its own. Even in high-pressure cluster environments such as Centaurus, M87 and Perseus, CO emission is detected only in a subset of the ionised filaments, indicating that additional local factors control molecular gas survival. Dust plays a critical role in H$_2$ formation.
	Filaments formed from uplifted, dust-rich interstellar medium (ISM) or preserved in younger, well-shielded structures are therefore more likely to host detectable molecular gas, whereas filaments formed predominantly through in-situ cooling of the dust-poor ICM are expected to be CO-faint despite being visible in H$\alpha$. Indeed, observations suggest that intermittent AGN outbursts can transport dust from galaxy centres to tens of kiloparsecs via buoyant outflows \citep[e.g.,][]{temi07}. Filaments are therefore likely to form from a combination of uplifted central gas and material that is freshly infalling or cooling from the ambient medium, which is expected to be relatively dust-poor. At the same time, dust exposed to the hot atmosphere can be efficiently sputtered, while AGN-driven turbulence and shocks may further disrupt grains, leading to an overall lower dust-to-gas ratio in these systems compared to $\sim 10^{-2}$ dust-to-gas ratios typical of star-forming disk galaxies.
	
	Nevertheless, observational evidence supports a close association between dust and molecular gas. Molecular clouds are spatially coincident with dust in the brightest group galaxies NGC~4636 and NGC~5846 \citep{temi18}, and in cluster filaments such as Perseus \citep{salome08}, Abell~1795 \citep{tamhane23}, and Centaurus \citep{olivares19}. In NGC~1316, regions with H$_2$/H {\sc i} $>1$ coincide with dust visible in HST imaging, while H {\sc i}-dominated regions lack visible dust and have lower column densities \citep{maccagni21}. Simulations further show that dust can survive or form in situ within multiphase outflows under favourable conditions \citep{qiu20,chen24}. Magnetic fields may further help prolong grain lifetimes, with recent MHD simulations showing that charged grains can couple to field lines, promoting magnetic draping around cool clumps and increasing the surviving dust mass by more than a factor of two compared to a non-magnetised ISM \citep[e.g.,][]{kirchschlager24}. These simulations suggest that variations in the local Alfv\'enic Mach number, such as those expected between group and cluster environments (see Figure~\ref{fig:B-turb}), alter the magnetic coupling of dust and change the efficiency of clustering and destruction. However, dust survival depends sensitively on time spent in the hot phase and declines with decreasing pressure and magnetic field strength, naturally producing lower dust-to-gas ratios at larger radii \citep{farber22,richie24}.
	
	Together, these results suggest that molecular gas survival in filaments is regulated by a combination of ambient pressure, turbulence, dust availability, magnetic support, and filament age. High pressure promotes gas retention, but dust determines where H$_2$ and CO can form. This framework explains the lack of molecular gas detection in some cluster filaments and the scarcity of extended molecular gas in groups such as NGC~5044, while remaining consistent with compact molecular reservoirs in galaxy centres. If magnetic fields are systematically stronger in cluster cores than in group environments, enhanced grain–field coupling and magnetic confinement could prolong dust survival and help maintain cold, dusty clumps embedded within ionised filaments, thereby contributing to both the larger radial extent of molecular gas and its closer spatial association with the warm phase in clusters. If magnetic fields are systematically stronger in cluster cores than in group environments, enhanced grain–field coupling and magnetic confinement could prolong dust survival and help maintain cold, dusty clumps embedded within ionised filaments, thereby contributing to both the larger radial extent of molecular gas and its closer spatial association with the warm phase in clusters.

	\section{SUMMARY AND CONCLUSION}
	
	We have presented new HST imaging of the ionised filaments in the brightest group galaxy NGC 5044 and compared their properties to filaments observed in cluster environments of M87, Centaurus and Perseus clusters. This is the first high-resolution view of filamentary structures in a galaxy group, allowing us to probe their widths, column densities, and stability. Our main conclusions are:
	
	\begin{itemize}
		\item The filaments in NGC 5044 extend several kiloparsecs from the centre and some of them remain coherent. Their widths span $\sim$50--120 pc, with some strands as narrow as those in clusters ($\sim$50 pc) but others significantly broader. This is consistent with the lower ambient pressure of the group environment, which allows filaments to be less tightly confined. We find that the filament width roughly scales with ambient pressure as FWHM $\propto P^{-0.4}$.
		
		\item Equipartition magnetic fields in filaments in NGC 5044 decline from $\sim$40 $\mu$G at the centre to $\sim$20 $\mu$G at 5 kpc, about 2--3 times lower than in cluster cores. Independent estimates from dynamical stability arguments against tidal forces imply much stronger radial fields of order $B_r \sim 10^2$ $\mu$G close to the centre, declining outward for dense gas component and likely lower for more diffuse components. These values are consistent with magnetic field strengths seen in recent MHD simulations of condensing gas, and can be explained by flux freezing in uplifted material and field draping around AGN-driven cavities. However, such high values may be difficult to reconcile with the relatively low fields inferred from RM, while turbulence and cosmic rays can provide complementary support.
		
		\item In both groups and clusters, the average filament surface densities lie well below the magnetic critical threshold, implying that filaments are gravitationally stable. Ultraviolet imaging shows no evidence of young stellar clusters with SFR upper limit of $<10^{-3}$ M$_\odot$ yr$^{-1}$, confirming suppressed star formation consistent with largely suppressed star formation in filaments across groups and clusters. Occasional star formation events may occur, for example, in regions compressed by jet--gas interactions.
		
		\item Extinction and infrared measurements in NGC~5044 ($A_V \sim 0.07$), imply a dust mass of $M_{\rm dust} = 3.2^{+18.4}_{-2} \times 10^5$~M$_\odot$ and a dust-to-gas ratio of $0.003^{+0.018}_{-0.002}$. A similar analysis for NGC~4696 gives $M_{\rm dust} = 4.1^{+2.2}_{-1.1} \times 10^5$~M$_\odot$ and a dust-to-gas ratio of $0.005^{+0.002}_{-0.001}$.
		
		\item The nuclear ionised gas within the Bondi radius exhibits a steeply declining H$\alpha$ surface brightness ($\propto r^{-1.07}$) and a low filling factor ($f \gtrsim 3 \times 10^{-3}$), implying an electron density profile $n_e \propto r^{-1.07}$ and an ionised gas mass of $< 2.8 \times 10^4$ M$_\odot$. This dense core is connected to extended, magnetically confined filaments that may channel gas toward the SMBH.
		
		\item Filaments are multiphase structures, consisting of cold gas embedded in diffuse atomic and ionised material, with molecular cloudlets present in some cases. Their survival is regulated by ambient pressure, dust shielding, and magnetic support. In low-pressure groups such as NGC 5044, weaker confinement, reduced magnetic shielding, and turbulent dispersal make molecular clumps rare, whereas in higher-pressure cluster cores, stronger pressure and better dust preservation allow molecular structures to persist and remain broadly co-spatial with the ionised filaments.
		
	\end{itemize}
	
	Overall, our results demonstrate that the filaments in galaxy groups share the same physical origin and stabilising mechanisms as those in clusters. Magnetic fields and AGN feedback act together to maintain filamentary structures and suppress star formation, even in systems with substantial molecular gas reservoirs. The observed contrast between group and cluster filaments is naturally explained if filament morphology is tied to ambient pressure and dust survival: groups, with lower thermal pressure, should host broader, more weakly confined structures, while the higher pressures in clusters favour narrower strands. This framework leads to a clear, testable prediction for future observations and simulations:\\
	(i) on average, filaments in higher-pressure cluster cores should be narrower than in groups,\\
	(ii) if magnetic tension provides the dominant confinement, filament-coincident RM structures should be detectable at $\leq 0.1$ kpc resolution, aligned with H$\alpha$ ridges, and\\
	(iii) group filaments should be dominated by diffuse ionised/H {\sc i} gas with only sporadic molecular clumps, whereas cluster filaments should show stronger co-spatiality of molecular and ionised phases.
	
	Extending the multiphase feedback paradigm to group environments therefore highlights both the universality of these processes and the opportunity to probe their signatures across the halo mass spectrum.
	
	\section*{Acknowledgement}	
		We thank Yuan Li for comments. We thank the anonymous referee for their comments that improved the paper.
		Based on observations with the NASA/ESA Hubble Space Telescope obtained from the Data Archive at the Space Telescope Science Institute, which is operated by the Association of Universities for Research in Astronomy, Incorporated, under NASA contract NAS5-26555. Support for Program number (15290) and (17037) was provided through grants from the STScI under NASA contract NAS5-26555. M.G. acknowledges support from the ERC Consolidator Grant \textit{BlackHoleWeather} (101086804). NW was supported by the GACR grant 21-13491X.
		This work has made use of data from the European Space Agency (ESA) mission
		{\it Gaia} (\url{https://www.cosmos.esa.int/gaia}), processed by the {\it Gaia}
		Data Processing and Analysis Consortium (DPAC,
		\url{https://www.cosmos.esa.int/web/gaia/dpac/consortium}). Funding for the DPAC
		has been provided by national institutions, in particular the institutions
		participating in the {\it Gaia} Multilateral Agreement. We made use of OpenAI's ChatGPT (GPT-5) to assist with language editing and polishing of the manuscript. All scientific content and conclusions are the responsibility of the authors.

	\section*{Data Availability}
	All of the data presented in this article were obtained from the Mikulski Archive for Space Telescopes (MAST) at the Space Telescope Science Institute. The specific observations analysed for NGC 5044 can be accessed via \url{https://dx.doi.org/10.17909/8x62-0x22}. The NGC 5044 mosaic images are available at MAST as a High Level Science Product via \url{https://doi.org/10.17909/ej9x-ah67}.
	
	\bibliographystyle{paslike_pasa}
	\bibliography{N5044-filaments}
	
	\appendix
	\section{F665N radial intensity profile}
	\label{f665n_profile}
\begin{figure}
		\centering
		\includegraphics[width=0.47\textwidth]{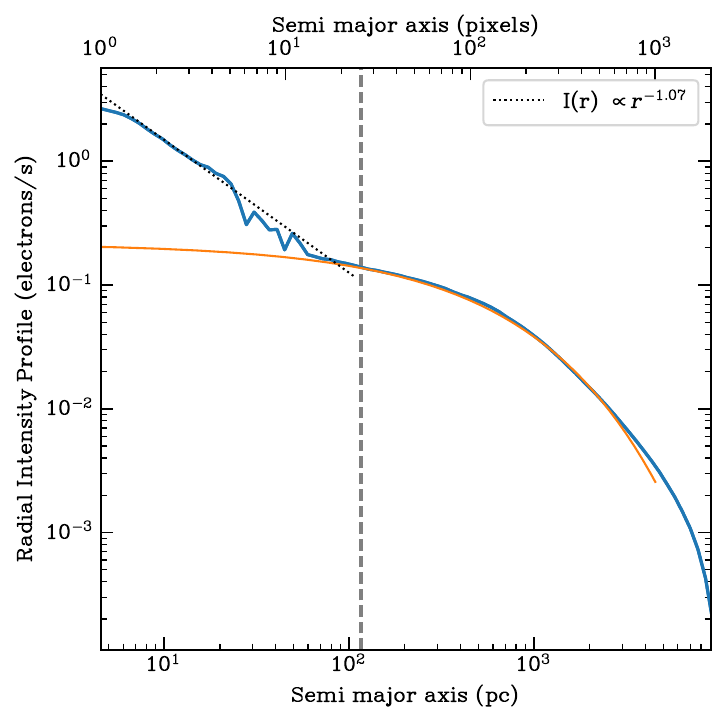}
		\caption{The radial intensity profile of the HST F665N image is shown in blue, with the S\'{e}rsic model fit to the galaxy continuum in orange, characterised by $r_{\rm eff} = 2.28$ kpc and $n = 1.609$. The plot shows the excess emission within the Bondi radius (dashed grey line), associated with the ionised gas core. The fit to the radial intensity profile inside the Bondi radius is shown as a dotted black line.}
		\label{fig:f665_sbprof}
	\end{figure}

	We removed the diffuse galactic emission from the F665N image using iterative isophotal fitting with \textsc{Photutils}, following the example notebooks\footnote{\url{https://github.com/astropy/photutils-datasets/tree/main/notebooks/isophote}}. The fitting procedure returns an \textsc{isophot} object containing the radial intensity profile and isophote shape parameters. In Figure~\ref{fig:f665_sbprof}, we present the derived radial intensity profile of NGC 5044 in the HST F665N filter. The galaxy continuum and the central core component are clearly distinguishable. To highlight this separation, we fitted a one-dimensional S\'{e}rsic model to the continuum profile, obtaining best-fit parameters of $r_{\rm eff} = 2.28$ kpc and $n = 1.609$. We also fitted the core with a powerlaw model of the form $I(r) \propto r^\alpha$, finding $\alpha = -1.07$. Small dips and wiggles in the core profile are likely due to dust absorption.

\end{document}